\documentclass[11pt,letterpaper]{article} 
\pdfoutput=1
\usepackage{jheppub}
\usepackage{bbm}
\usepackage{mathrsfs}
\usepackage{slashed}
\usepackage{caption}
\usepackage{epstopdf}
\usepackage[normalem]{ulem}
\usepackage[bottom]{footmisc}
\usepackage{subcaption}
\usepackage{bbold}
\usepackage{titlesec}
\usepackage{threeparttable}
\usepackage{booktabs}
\usepackage{changepage}
\usepackage[utf8]{inputenc}
\usepackage{dsfont}

\usepackage{grffile}
 
\usepackage{graphicx}  
\usepackage{dcolumn}   
\usepackage{bm}        
\usepackage{amssymb}   
\usepackage{setspace}
\usepackage{amsmath, amssymb, setspace}
\usepackage{array}
\usepackage{booktabs}
\usepackage{caption}
\usepackage{indentfirst}
\usepackage{float}
\usepackage{lmodern}
\usepackage{multirow}
\usepackage{soul}
\usepackage[normalem]{ulem}

\usepackage{braket}
\usepackage{comment}

\usepackage[draft]{pgf}

\usepackage{adjustbox} 
\newlength{\myimageoversize}
\newsavebox{\myimage}

\usepackage[most]{tcolorbox}

\usepackage{empheq}

\tcbset{colback=white!10!white, colframe=black!50!black, 
        highlight math style= {enhanced, 
            colframe=black,colback=white!10!white,boxsep=0pt}
        }
\titleformat{\subsection}
 {\normalfont\fontsize{12}{17}\itshape}{\thesubsubsection}{1em}{}
 
\title{\huge{Multiple Populations of Same Sterile Neutrino as Dark Matter}}

\author[c,d,a]{Muping Chen,}
\author[a]{Graciela B. Gelmini,}
\author[b]{Philip Lu,}
\author[c,d,e,f]{and Volodymyr Takhistov} 

\affiliation[a]{Department of Physics and Astronomy, UCLA,
475 Portola Plaza, Los Angeles, CA 90095, USA}
\affiliation[b]{Center for Theoretical Physics, Department of Physics and Astronomy, Seoul National University, Seoul 08826, Korea}
\affiliation[c]{International Center for Quantum-field Measurement Systems for Studies of the Universe and Particles (QUP,WPI), High Energy Accelerator Research Organization (KEK), Oho 1-1, Tsukuba, Ibaraki 305-0801, Japan}
\affiliation[d]{Theory Center, Institute of Particle and Nuclear Studies (IPNS),
High Energy Accelerator Research Organization (KEK), Tsukuba 305-0801, Japan} 
\affiliation[e]{Graduate University for Advanced Studies (SOKENDAI), \\
1-1 Oho, Tsukuba, Ibaraki 305-0801, Japan}
\affiliation[f]{Kavli Institute for the Physics and Mathematics of the Universe (WPI), Chiba 277-8583, Japan}

\emailAdd{mpchen@physics.ucla.edu}
\emailAdd{gelmini@physics.ucla.edu}
\emailAdd{philiplu11@gmail.com}
\emailAdd{vtakhist@post.kek.jp}

\abstract{Sterile neutrinos produced in the early Universe that mix with active neutrinos of the Standard Model are typically considered to 
consist of a single population resulting from one dominant production mechanism. We show that the same sterile neutrino species can naturally emerge with  multiple population components, yielding a multi-modal relic momentum spectrum. We consider this with four distinct production scenarios: active-sterile non-resonant oscillations following resonant oscillations 
in the presence of a primordial lepton asymmetry, gravitational production through sterile neutrinogenesis from populations of evaporating primordial black holes, and heavy singlet Higgs or inflaton decays combined with non-resonant active-sterile oscillations or neutrinogenesis.  
We identify sterile neutrino mass ranges where a colder and a hotter population can be present with similar contributions and can also contribute non-negligibly to the dark matter relic abundance.  We discuss some potential consequences of such a multi-population framework.  
}

\begin{document}
\preprint{KEK-QUP-2025-0016, KEK-TH-2738}

 \maketitle
\flushbottom

\section{Introduction}

Sterile or right–handed neutrinos, $\nu_s$, provide one of the simplest extensions of the Standard Model (SM). They can compose all or part of the dark matter (DM) that constitutes about $85\%$ of all matter in the Universe, has thus far been detected only through its gravitational interactions, and whose nature remains one of the most important open problems in physics, astrophysics and cosmology~(e.g.~\cite{Gelmini:2015zpa,Bertone:2016nfn}).
They are also simultaneously motivated by a variety of outstanding problems in physics, such as the origin of the small masses of active neutrinos and of the matter-antimatter asymmetry in the Universe.

These neutrinos are fermions that do not directly couple to the SM weak gauge bosons, and in the simplest models
only couple to the SM through their mixing with active neutrinos $\nu_\alpha$ ($\alpha=e,\mu,\tau$). 
Sterile neutrinos can naturally generate the small 
masses of active neutrinos observed in oscillation experiments (e.g.~\cite{Super-Kamiokande:1998kpq}), through a see-saw mechanism~\cite{Yanagida:1979as,Gell-Mann:1979vob} in minimal models. For masses $m_s$ in the $\mathcal{O}(\text{keV})$ range, the lightest sterile species constitutes an excellent candidate for warm DM,
and can appear in variety of frameworks (see e.g. ~\cite{Boyarsky:2018tvu} for a review).  Radiative decays of $\mathcal{O}(\text{keV})$-mass sterile neutrinos 
that are part of the DM would result in  observable X-ray signatures\footnote{A putative line at $3.5$~keV observed in clusters and galaxies~\cite{Bulbul:2014sua,Boyarsky:2014jta} has been linked to a $m_s\simeq7$~keV sterile neutrino, although this interpretation has been challenged (e.g.~\cite{Jeltema:2014qfa,Dessert:2023fen}).},
which places  stringent limits  
on their mixing with active neutrinos 
dependent on their DM abundance~\cite{Ng:2019gch,Perez:2016tcq,Neronov:2016wdd}. These decays 
will be further tested with future observations by instruments such as X-Ray Imaging and Spectroscopy Mission (XRISM)~\cite{Dessert:2023vyl,Zhou:2024sto}.
Throughout this work, we consider a single sterile species that predominantly mixes with the electron flavor active neutrino\footnote{Our analysis can readily accommodate mixing with other flavors.} $\nu_e$ for concreteness.

The cosmological role of the DM in the formation of the large scale structure of the Universe 
is governed by its primordial momentum distribution.  
Large scale structure observations imply that the bulk of the DM must be either non-relativistic (``cold" DM, CDM), or becoming non-relativistic (``warm" DM, WDM) when the  temperature of the Universe was of $\mathcal{O}$(keV). Only a minor relativistic component of DM (``hot" DM, HDM)  is allowed to be present at that moment.
Potential tensions related to small scale structure in the Universe could be suggestive of going beyond the prevalent CDM paradigm~\cite{deBlok:2009sp,Boylan-Kolchin:2011qkt}. 
Free-streaming of WDM 
suppresses the matter power spectrum below characteristic scales constrained by  
Lyman-$\alpha$ forest
observations~\cite{Viel:2005qj,Baur:2017stq,Garcia-Gallego:2025kiw}, by substructure in strong gravitational lenses~\cite{Gilman:2019nap,Zelko:2022tgf}, and by the abundance of dwarf satellites, among other data. Sterile neutrinos consisting of a colder and a warmer 
population admixture therefore offer an enticing scenario to explore.
 
Typically production mechanisms 
are considered to generate a single relic sterile neutrino population. Production via non-resonant active-sterile oscillations (or Dodelson-Widrow, DW, mechanism)~\cite{Dodelson:1993je}, and resonant active-sterile oscillations in the presence of a large primordial lepton asymmetry (or Shi-Fuller mechanism)~\cite{Shi:1998km} depend on the active-sterile mixing.
Alternatively, sterile neutrinos could be gravitationally  produced entirely independently of this mixing in the Hawking evaporation of primordial black holes (PBHs), or ``PBH sterile neutrinogenesis"~\cite{Chen:2023lnj,Chen:2023tzd}\footnote{Gravitational cosmological pair production of sterile neutrinos during Universe expansion is also possible, but is strongly suppressed for light sterile neutrinos. On the other hand, post-inflationary gravitational particle production from inflaton oscillations relies on unknown quantum-gravity–induced operators~\cite{Koutroulis:2023fgp}.}. 
Through PBH neutrinogenesis, sterile neutrinos are produced with exceedingly large initial momentum 
corresponding to the PBH Hawking temperature, and a spectrum that is distinct from other processes.  
Yet another possibility is sterile neutrino production in decays of heavy particles, such as the inflaton or a singlet Higgs (e.g.~\cite{Shaposhnikov:2006xi,Kusenko:2006rh, Petraki:2007gq,Merle:2013wta,Merle:2015oja,Abazajian:2019ejt}), a mechanism that can also produce a significant population of sterile neutrinos independent of  
their mixing with active neutrinos, relying instead on other particle interactions.

In this work, we explore the possibility  
that more than one of such sterile neutrino production mechanisms could occur at different cosmological epochs, leading to sterile neutrino populations with different characteristics, in particular distinct relic momentum distributions.
Scenarios with multiple sterile neutrino populations can thus offer novel insights into potential production mechanisms as well as intriguing sterile neutrino features and observables.
Hence, through mechanisms operating at disjoint cosmological epochs, a single $\nu_s$ species can inherit multiple momentum components, a colder one and a hotter one, each contributing to the relic density.  
Some related  
considerations were presented in earlier works (e.g.~\cite{Boyarsky:2008mt,Merle:2014xpa, Merle:2015oja,Merle:2015vzu,Abazajian:2019ejt}). Here, we undertake a detailed quantitative analysis of multiple populations of sterile neutrino DM and illustrate several conceptually distinct scenarios they can arise in.

This paper is organized as follows. 
In Sec.~\ref{sec:nonresprod}, we introduce the Boltzmann evolution
framework for sterile neutrino phase-space, and  
discuss sterile neutrino production from non-resonant and resonant active-sterile neutrino flavor oscillations. We point out the possible 
appearance of two sterile neutrino populations when considerable non-resonant production follows resonant production.
In Sec.~\ref{sec:neutrinogenesis}, we consider gravitational sterile neutrino production from PBH neutrinogenesis together with flavor oscillations, and from multiple PBH populations. 
In Sec.~\ref{sec:heavydecay}, we examine two sterile neutrino populations appearing from decays of heavy singlet Higgs together with flavor oscillations.
For each scenario, we track the complete momentum 
distribution. In Sec.~\ref{sec:limits} we confront observations  and outline signatures that can test such mixed-population framework.  
We summarize our main findings in Sec.~\ref{sec:conclusion}.

\section{Production via Resonant and Non-Resonant Active-Sterile Oscillations}
\label{sec:nonresprod}

Due to their mixing with active neutrinos, sterile neutrinos will inevitably be produced through active-sterile flavor oscillations even when other production mechanisms are simultaneously active. This mechanism, extensively explored for generating DM abundance, can be classified into non-resonant (Dodelson-Widrow) and resonant (Shi-Fuller) production regimes depending on the primordial lepton asymmetry in the early Universe. However, in cosmological scenarios, both resonant and non-resonant production can coexist and contribute to the sterile neutrino phase-space distribution. In this section, we systematically analyze the evolution of sterile neutrino production from oscillations, explicitly tracking the contributions from resonant and non-resonant channels over cosmic time. We follow the time-dependent depletion of lepton asymmetry, include antineutrino effects, and solve the relevant Boltzmann equations without assuming a fixed temperature production epoch. This allows us to construct  full sterile neutrino momentum distribution and to explicitly identify and separate the resulting colder and warmer components.

\subsection{Boltzmann formalism}
\label{sec: boltzmanneq}

To accurately model sterile neutrino production via oscillations, it is essential to solve the Boltzmann equation governing their momentum distribution evolution. This framework captures the interplay between cosmological expansion, active-sterile conversion rates, and the evolution of the primordial lepton asymmetry. A complete quantum treatment would require solving the full quantum kinetic equations~\cite{Stodolsky:1986dx,McKellar:1992ja}, but for the regimes relevant here we adopt the well-established relaxation-time approximation that reduces this to a semiclassical Boltzmann equation. This approach has been shown to reproduce the full quantum results with high accuracy in most cases~\cite{Johns:2019hjl}. We also assume conservative scattering, where an active neutrino converts into a sterile neutrino of the same momentum, a simplification that introduces only negligible errors~\cite{Dolgov:2000ew}. 
 
The resulting Boltzmann equation for the sterile-neutrino momentum  
distribution function $f_{\nu_s}(p,t)$ takes the form (see e.g. \cite{Abazajian:2001nj, Kishimoto_2008} and references therein)
\begin{eqnarray}
\label{eq:boltzmannp}
    \frac{\partial}{\partial t}f_{\nu_s}(p,t)-Hp\frac{\partial}{\partial p}f_{\nu_s}(p,t)\simeq \Gamma(\nu_\alpha\rightarrow\nu_s;p,t)[f_{\nu_\alpha}(p,t)-f_{\nu_s}(p,t)]~,
\end{eqnarray}
where $p$ is momentum, $t$ is time and $\Gamma(\nu_\alpha \rightarrow \nu_s; p, t)$ denotes the momentum- and time-dependent active-to-sterile conversion rate. We have neglected Fermi-blocking effects, as the relevant phase-space occupation numbers remain well below unity. The distribution function $f_{\bar{\nu_s}}(p,t)$ for sterile antineutrinos satisfies an analogous equation, with $(\nu_\alpha, \nu_s)$ replaced by $(\bar{\nu}_\alpha, \bar{\nu}_s)$.

A technical advantage in our treatment is the use of the comoving momentum variable $\epsilon_g$
and generalized temperature $T_g$, which are invariant under cosmological expansion. This choice simplifies the calculations, especially when tracking production from different sources at different epochs. In particular, it allows the time-temperature relation to retain the same functional form while correctly incorporating changes in the number of relativistic degrees of freedom. This also simplifies the Boltzmann equation when rewritten in terms of temperature instead of time. 

\begin{figure}[t]
    \centering
    \includegraphics[width=0.49\linewidth]{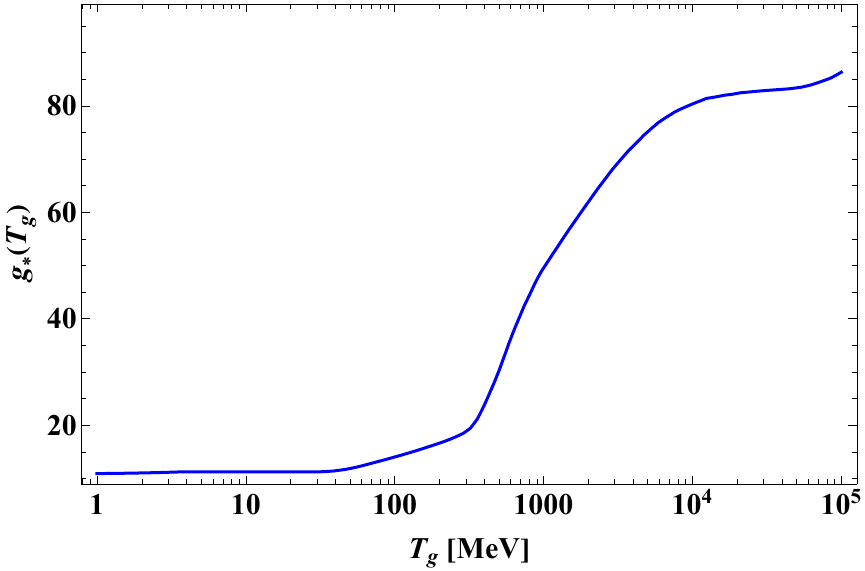}
    \caption{The effective number of relativistic degrees of freedom for energy and entropy densities $g_*=g_{s*}$ as a function of scaled temperature $T_g$, with $g_*(T)$ taken from Ref.~\cite{Borsanyi:2016ksw}.} 
    \label{fig:gstar}
\end{figure}

 In particular, we   can simplify Eq.~\eqref{eq:boltzmannp} by introducing the dimensionless temperature-scaled momentum variable $\epsilon = p/T$, which remains constant with time for relativistic neutrinos as long as 
the entropy  number of  degrees of freedom $g_{s*}$ is constant.
This approximation is often valid since sterile-neutrino production typically occurs over a narrow temperature range  in which $g_{s*}$ does not change. However, when considering production at multiple epochs, as we consider in our scenario,
this assumption no longer holds. The momentum of relativistic particles redshifts as $p \propto 1/a$, where $a$ is the cosmic scale factor. Entropy conservation per comoving volume implies that the product $T g_{s*}^{1/3} a$ remains constant, which from the definition of the Hubble expansion rate $H= \dot{a}/a$ implies   
\begin{eqnarray}
    \label{eq:dTg1/3dt}
    \frac{d(Tg_{s*}^{1/3})}{dt}=-Tg_{s*}^{1/3} H~,
\end{eqnarray}
and also that $\epsilon \propto g_{s*}^{1/3}$, since $T \propto 1/(g_{s*}^{1/3}a)$. Thus the combination $\epsilon g_{s*}^{-1/3}$ is conserved, and we conveniently define
\begin{eqnarray}
    \epsilon_g=\epsilon g^{-1/3}_{s*},~~~~~ T_g= Tg^{1/3}_{s*}~.
\end{eqnarray}
Thus, $T_g$ fulfills the same equation fulfilled by $T$ when $g_{s*}$ is constant, namely
\begin{eqnarray}
    \label{eq:dTgdt}
    \frac{dT_g}{dt}=-T_g H~.
\end{eqnarray} 

In the temperature range relevant to our analysis, above a few MeV, the effective number of entropy and energy degrees of freedom are approximately equal, $g_{s*} \simeq g_{*}$. Hence, we designate both with  $g_*$. Fig.~\ref{fig:gstar} shows 
$g_*$  as a function of $T_g$  that we employ 
in our numerical computations of sterile-neutrino production via neutrino flavor oscillations.

 Eq.~\eqref{eq:dTgdt} implies
\begin{eqnarray}
\label{eq:dfdTg}
  \left(\frac{\partial f_{\nu_s}}{\partial t}\right)_p= -HT_g \left(\frac{\partial f_{\nu_s}}{\partial T_g}\right)_{p}~,
\end{eqnarray}
and since $p= \epsilon_g T_g$ and $\epsilon_g$ is constant in time,
\begin{eqnarray}
\label{eq:dfdT@eps-1}
    \left(\frac{\partial f_{\nu_s}}{\partial T_g}\right)_{\epsilon_g}=\left(\frac{\partial f_{\nu_s}}{\partial T_g}\right)_{p}+\frac{p}{T_g}\left(\frac{\partial f_{\nu_s}}{\partial p}\right)_{T_g}~.
\end{eqnarray}
Thus, the Boltzmann equation in Eq.~\eqref{eq:boltzmannp} can be written as  
\begin{equation}
    \label{eq:boltz-eg}
    \left(\frac{\partial}{\partial T_g}f_{\nu_s}(\epsilon_g,T_g)\right)_{\epsilon_g}\simeq-\frac{1}{HT_g}\Gamma(\nu_\alpha\rightarrow\nu_s;\epsilon_g,T_g)[f_{\nu_\alpha}(\epsilon_g,T_g)-f_{\nu_s}(\epsilon_g,T_g)]~.
\end{equation}
Since we consistently find that $f_{\nu_s} \ll f_{\nu_\alpha}$ throughout the relevant temperature range, we neglect the back-reaction term proportional to $f_{\nu_s}$ in Eq.~\eqref{eq:boltz-eg},
\begin{equation}
  \label{eq:boltz-eg-approx}
    \frac{\partial}{\partial T_g}f_{\nu_s}(\epsilon_g,T_g)\simeq -\frac{1}{HT_g}\Gamma(\nu_\alpha\rightarrow\nu_s;\epsilon_g,T_g)f_{\nu_\alpha}(\epsilon_g,T_g)~.
\end{equation}
Active neutrinos and antineutrinos are in thermal equilibrium over the temperature range of interest, and thus follow Fermi-Dirac distributions,  
\begin{eqnarray}
    f_{\nu_\alpha}(\epsilon_g,T_g) &=&\dfrac{1}{ \exp\{\epsilon_g [g_*(T_g)]^{1/3}-\xi(T_g)\}+1 }
    \\
    f_{\bar{\nu}_\alpha}(\epsilon_g,T_g)&=&\dfrac{1}{\exp\{\epsilon_g [g_*(T_g)]^{1/3}+\xi(T_g)\}+1}~,
\end{eqnarray}
where  the chemical potential $\xi(T_g)$ is related to the lepton asymmetry $L_\alpha$ in the $\alpha$-flavor active neutrino species, $\xi(T_g) \simeq (12\zeta(3)/\pi^2)(g_*(T_g))^{-1/3}L_\alpha(T_g)$. Here $L_\alpha$ is defined as the net lepton asymmetry per photon number density
\begin{eqnarray}
\label{eq:leptonnum}
    L_\alpha(t) = \frac{n_{\nu_\alpha}(t) -n_{\bar{\nu}_\alpha}(t)}{n_\gamma(t)}~,
\end{eqnarray}
where $n_{\nu_\alpha}(t)$ and $n_{\bar{\nu}_\alpha}(t)$ are the active neutrino and antineutrino number densities, respectively, and $n_\gamma(t)=2\zeta(3)T_g^3(t)[g_*(t)]^{-1}/\pi^2$ is the photon number density. 
 
The conversion rate in the Boltzmann equation is the product of half of the total collision rate $\Gamma_\alpha$ and the average probability that an active neutrino has transformed into a sterile neutrino~(e.g.~\cite{Abazajian:2001nj}) 
\begin{eqnarray}
\label{eq:totalrate}
\Gamma(\nu_\alpha\rightarrow\nu_s;\epsilon_g,T_g)
\simeq \frac{\Gamma_\alpha}{2}\langle P(\nu_\alpha\rightarrow\nu_s;\epsilon_g,T_g) \rangle \simeq \dfrac{\Gamma_{\alpha}}{4}\dfrac{\sin^2 2\theta_m}{1+(\Gamma_\alpha \ell_m/2)^2}~.
\end{eqnarray}
where we have used in the last equality the expression for the average active-sterile conversion probability $\langle P(\nu_\alpha\rightarrow\nu_s;\epsilon_g,T_g) \rangle$, and   $\Gamma_\alpha$ is the $\alpha$ active neutrino  collision rate,  given below in Eq.~\eqref{eq:collisionrate}. 
Here,  $\sin\theta_m$ is the in-medium mixing 
\begin{eqnarray}
\label{eq:sin2thetam}
    \sin^2 2\theta_m(\epsilon_g,T_g)=\dfrac{{\sin^2 2\theta}}{ \sin^2 2\theta+\left(\cos 2\theta-\dfrac{2\epsilon_g T_g}{m_s^2}\left(V_T(\epsilon_g,T_g)+V_D(T_g)\right)\right)^2 }
\end{eqnarray}
and  
\begin{eqnarray}
\label{eq:osclength}
\ell_m(\epsilon_g,T_g)=\dfrac{1}{\left[ \left( \dfrac{m_s^2}{2\epsilon_g T_g}\sin 2\theta\right)^2+\left( \dfrac{m_s^2}{2\epsilon_g T_g}\cos2\theta -V_T(\epsilon_g,T_g)-V_D(T_g)\right)^2 \right]^{1/2}}
\end{eqnarray}
is the in-medium oscillation length.  These quantities are defined in terms of the thermal and density potentials, $V_T$ and $V_D$. 

The thermal potential $V_T(\epsilon_g,T_g)$ 
\begin{eqnarray}
\label{eq:VT}
    V_T(\epsilon_g,T_g)=-r_\alpha G_F^2\epsilon_g T^5_g [g_*(T_g)]^{-4/3}~,
\end{eqnarray}
accounts for finite-temperature effects and the neutrino forward scattering.
Here, $G_F$ is the Fermi constant  and $r_\alpha$~\cite{Abazajian:2001nj, Abazajian:2004aj, Kishimoto_2008} is a dimensionless coefficient that denotes the contribution from charged leptons of the same flavor present in the thermal bath. The value of $r_\alpha$ generally varies with temperature and neutrino flavor. For electron flavor, $\alpha = e$,  
 it is a good approximation to  adopt a constant value of $r_e = 79.34$ across all temperatures considered.

In the early Universe, the finite density potential $V_D(T_g)$ is 
dominated by the lepton asymmetries  if they are much larger than the baryon asymmetry,   
\begin{eqnarray}
\label{eq:VD}  
V_D(T_g)=\frac{2\sqrt{2}\zeta(3)}{\pi^2}G_FT^3_gg_*^{-1}(T_g)\left(\mathcal{L}_\alpha(T_g)\pm\frac{\eta_B}{4}\right),
\end{eqnarray} 
where the $+$ sign applies to $\nu_e$, and the $-$ sign to $\nu_\mu$ and $\nu_\tau$. Here $\eta_B$ is the baryon-to-photon ratio and $\mathcal{L}_\alpha(T_g)$ is the 
effective net lepton number relevant for forward scattering, defined as
\begin{eqnarray}
\label{eq:curlL}  \mathcal{L}_\alpha(T_g)= 2L_\alpha(T_g)+\sum_{\beta\neq\alpha}L_\beta(T_g)~.
\end{eqnarray}
where $\alpha$ and $\beta$ denote active neutrino flavors. 

We consider large initial  neutrino  lepton asymmetries $\mathcal{L}_0$, much larger than the baryon to photon ratio, $\mathcal{L}_0\gg \eta_B$. Thus, contributions to the forward scattering potential from neutrino–baryon and neutrino–charged lepton interactions can be neglected, and $V_D$ simplifies to
\begin{eqnarray}
    V_D(T_g)\simeq \frac{2\sqrt{2}\zeta(3)}{\pi^2}G_F\mathcal{L}_\alpha(T_g)T^3_gg_*^{-1}(T_g)~.
\end{eqnarray}

Combining Eq.~\eqref{eq:totalrate}, Eq.~\eqref{eq:sin2thetam}, and Eq.~\eqref{eq:osclength}, we obtain the total conversion rate
\begin{eqnarray}
\label{eq:totalrate-nu}
\Gamma(\nu_\alpha\rightarrow\nu_s)
\simeq \dfrac{\Gamma_\alpha}{4}~\dfrac{\sin^2 2\theta}{\sin^2 2\theta+D^2+\left(\cos 2\theta-\dfrac{2\epsilon_g T_g}{m_s^2}(V_T+V_D)\right)^2}~,
\end{eqnarray}
and similarly
\begin{eqnarray}
\label{eq:totalrate-nubar}
\Gamma(\bar{\nu}_\alpha\rightarrow\bar{\nu}_s)
\simeq \frac{\bar{\Gamma}_\alpha}{4}~\dfrac{\sin^2 2\theta}{\sin^2 2\theta+\bar{D}^2+\left(\cos 2\theta-\dfrac{2\epsilon_g T_g}{m_s^2}(V_T-V_D)\right)^2}~
\end{eqnarray}
for antineutrinos.
Here $D=\Gamma_\alpha\epsilon_gT_g/m_s^2$   and $\bar{D}=\bar{\Gamma}_\alpha\epsilon_gT_g/m_s^2$ are the quantum damping factors (or decoherence rates) that quantify the suppression of active–sterile oscillations due to frequent interactions of active neutrinos with the thermal plasma~\cite{Stodolsky:1986dx,Abazajian:2001nj,PhysRevD.59.113001}. These interactions effectively ``measure'' the flavor state of the neutrino, leading to decoherence via the quantum Zeno effect and thus inhibiting coherent oscillations. The damping terms enter the oscillation probability and suppress sterile neutrino production in the strongly interacting regime.

To linear order in $L_\alpha$, the active neutrino collision rates are
\begin{equation}
\begin{aligned}
     \label{eq:collisionrate}
 &\Gamma_\alpha (\epsilon_g,T_g) \simeq y_\alpha(T_g)G_F^2\epsilon_gT^5_gg^{-4/3}_*(T_g)(1-z_\alpha L_\alpha)\\
    &\bar{\Gamma}_\alpha (\epsilon_g,T_g) \simeq y_\alpha(T_g)G_F^2\epsilon_gT^5_gg^{-4/3}_*(T_g)(1+z_\alpha L_\alpha)  ~,
    \end{aligned}
\end{equation}
where $z_e\simeq 0.1$ for $\alpha=e$ and $z_{\mu,\tau}\simeq0.04$ for $\alpha=\mu,\tau$~\cite{PhysRevD.59.113001}. 
Considering observational upper limits of $L_{\alpha} \ll 1$, one has
\begin{eqnarray}
\Gamma_\alpha (\epsilon_g,T_g) \simeq \bar{\Gamma}_\alpha (\epsilon_g,T_g)  \simeq y_\alpha(T_g)G_F^2\epsilon_gT^5_gg^{-4/3}_*(T_g)~,
\end{eqnarray}
This total collision rate  is common to both the $\nu_\alpha\rightarrow \nu_s$ and $\bar{\nu}_\alpha\rightarrow \bar{\nu}_s$ processes. 

The coefficient $y_\alpha(T_g)$ accounts for the number of relativistic particles carrying weak charge in the thermal bath and therefore modulates the total interaction rate. At higher temperatures, additional species such as charged leptons and quarks contribute, enhancing the rate through forward scattering.
For the electron neutrinos we are considering, we adopt the function $y_e(T_g)$
from Ref.~\cite{Kasai:2024diy}. This function was fitted to the numerical results given in Ref.~\cite{Asaka:2006nq} for fixed scaled momentum $\epsilon=3$, which introduces errors of at most $\lesssim20\%$. 
The value of $y_e(T_g)$  
is close to unity for $T_g < 45$ MeV and  grows with $T_g$ to reach the value 8 at $T_g\sim 4$ GeV. 

Examining Eqs.~\eqref{eq:totalrate-nu} to~\eqref{eq:collisionrate},
we note that $\Gamma(\bar{\nu}_\alpha\rightarrow\bar{\nu}_s)$ differs from $\Gamma(\nu_\alpha\rightarrow\nu_s)$ only by the sign 
of the terms linear in $L_\alpha$. 
This sign difference leads to significant distinctions between the sterile neutrino conversion rates for neutrinos and antineutrinos.
For large  positive values of $L_\alpha$ where $V_D > |V_T|$, with $V_T$ being negative, the total potential $V=V_T+V_D$ for neutrinos can become positive. This leads to a resonant enhancement at a specific value of $\epsilon$ corresponding to the vanishing of the term in parentheses in Eq.~\eqref{eq:totalrate-nu}. We denote this condition as  $F_{\rm res} = 0$, where
\begin{eqnarray}
    \label{eq:Fres}
    \begin{aligned}
   F_{\rm res}(\epsilon_g,T)&= 
   \cos2\theta-\frac{2\epsilon_g T_g}{m_s^2}\left[ V_T(\epsilon_g,T_g)+V_D(\mathcal{L},T_g) \right]\\
   &=~ \cos 2\theta-\frac{4\sqrt{2}\zeta(3)G_F}{\pi^2 m_s^2}\mathcal{L}_\alpha(T)\epsilon_g  T^4_g g_*^{-1}+\frac{2r_\alpha G_F^2}{m_s^2}\epsilon_g^2  T^6_g g_*^{-4/3}~.
   \end{aligned}
\end{eqnarray} 
For positive values of $L_\alpha$, the total potential is negative 
for antineutrinos,  and  
there is no resonance for $\bar{\nu}_{\alpha} \rightarrow \bar{\nu}_s$. Furthermore, the sign difference in the $V_D$ term ensures that $\Gamma(\nu_\alpha\rightarrow\nu_s)>\Gamma(\bar{\nu}_\alpha\rightarrow\bar{\nu}_s)$, leading to a net depletion of the lepton number. As shown in Eq.~\eqref{eq:dLdtv2}, the time derivative of the lepton asymmetry satisfies $\dot{L}_\alpha (t) < 0$ indicating that   
$L_\alpha$ is depleted over time. 
This lepton number depletion has two important consequences. First, there can exist a termination time $t_{\rm end}$ after which the resonance condition is no longer satisfied, halting resonant sterile neutrino production. Second, the quantum damping factor decreases with time as $L_{\alpha}$ is depleted. This reduction in damping can enhance the sterile neutrino production rate at later times, particularly in the non-resonant regime.

In the limit $\mathcal{L}_\alpha\xrightarrow[]{} 0$, the resonance condition $F_{\rm res}(\epsilon_g,T_g)=0$ cannot be satisfied, and sterile neutrinos are produced solely through non-resonant (DW) production. In this case, the active–sterile conversion rates are identical for neutrinos and antineutrinos, $\Gamma(\nu_\alpha\rightarrow\nu_s)=\Gamma(\bar{\nu}_\alpha\rightarrow\bar{\nu}_s)$. The time derivative of the lepton number vanishes, $\dot{L}_\alpha=0$ (see Eq.~\eqref{eq:dLdt}), and
the non-resonant condition  $\mathcal{L}_\alpha \simeq 0$ is maintained throughout evolution.

At present time, all sterile neutrinos we consider  
are non-relativistic and their energy density is  $\rho_{\nu_s,0}=m_s  n_{\nu_s,0}$. 
This density can be expressed as a fraction of the DM density $f_{\rm s,DM}=\rho_{\nu_s,0}/\rho_{\rm DM}$, where we use $\rho_{\rm DM}=1.26\times10^{-6}\textrm{ GeV}/\textrm{cm}^3$~\cite{PhysRevD.98.030001} and the decoupling temperature $T_{\rm dc} \simeq 3 \textrm{ MeV}$, 
yielding 
\begin{align}
\label{eq:fsdm}
    f_{\rm s,DM}&=\frac{2m_sn_{\nu_\alpha,0}g_*(T_{\rm dc})}{3\zeta(3)\rho_{\rm DM}}
 \int d\epsilon_{g}   \epsilon_{g}^2 [f_{\nu_s}(\epsilon_{g})+f_{\bar{\nu}_s}(\epsilon_g)]\notag\\
&\simeq 270\left(\frac{m_s}{\rm keV} \right)\left( \frac{g_*(T_{\rm dc})}{10.75} \right)  \int d\epsilon_{g}   \epsilon_{g}^2 [f_{\nu_s}(\epsilon_{g})+f_{\bar{\nu}_s}(\epsilon_g)]~,
\end{align} 
which must satisfy $f_{\rm s,DM}\leq1$ to not overclose the Universe.

For negligible initial lepton asymmetry, sterile neutrinos are produced via non-resonant DW production. 
In this case, the sterile neutrino production rate has a sharp maximum at a temperature~\cite{Abazajian:2001nj} 
\begin{equation}
\label{eq:Tmax}
    T_{\rm max}\simeq 133 \textrm{ MeV}\left(\frac{m_s}{1\textrm{ keV}}\right)^{1/3}~.
\end{equation}
Taking the approximation of a constant $g_*(T) \simeq g_*(T_{\rm max})$, the sterile neutrino density from DW production can be analytically approximated as~\cite{Gelmini:2019wfp} (see Sec.~\ref{sec:nonresprod} for details)
\begin{equation}
\label{eq:fsdmnonresapprox}
    f_{\rm s,DM} \simeq 6.6\times10^{-4}\left(\frac{\sin^2 2\theta}{10^{-10}}\right)\left(\frac{m_s}{\textrm{keV}}\right)^2 \left(\frac{y_\alpha}{3}\right)\left(\frac{g_*(T_{\rm max})}{30}\right)^{-3/2}~.
\end{equation}
In our numerical calculation of non-resonant production, the lepton number is never small enough to realize pure DW production. Therefore, the temperature at which sterile neutrino production peaks and the resulting DM fraction 
differ from estimates given in Eq.~\eqref{eq:Tmax} and~Eq.~\eqref{eq:fsdmnonresapprox}.

In the following, we present our results in terms of the physical scaled momentum $\epsilon(t)$ at a chosen reference time $t=t_{\rm ref}$, which we call $\epsilon_{\rm ref}$. At the reference time $t=t_{\rm ref}$, the scaled momentum is defined as  
\begin{equation}
    \epsilon_{\rm ref}=\epsilon_g  g_{*,\rm ref}^{1/3}~.
\end{equation}
We choose $\epsilon_{\rm ref}$ as the value of $\epsilon(t)$  at $T= 3$ MeV, approximately corresponding to the epoch of active neutrino decoupling.  The physical scaled momentum at any time $t$ can then be written in terms of the reference momentum,
\begin{equation}
    \epsilon(t)=\epsilon_g g_*^{1/3}(t)=\epsilon_{\rm ref}\left(\dfrac{{g_*(t)}}{{g_{\rm *,ref}}}\right)^{1/3}.
\end{equation}

In our analysis we consider a nonzero lepton asymmetry only in the electron sector, setting $L_{\mu,\tau} = 0$, so that $\mathcal{L}_e (T_g)=2L_e(T_g)$. Without  taking into account flavor oscillations among active neutrinos, which happens at $T< 10$ MeV, constraints from Big Bang Nucleosynthesis (BBN) and the Cosmic Microwave Background (CMB) restrict the lepton asymmetry in electron neutrinos to $|L_e| \lesssim \mathcal{O}(10^{-3})$ (and impose  weaker bounds for the other flavors $|L_{\mu,\tau}|\lesssim \mathcal{O}(10^{-1})$)~\cite{Mangano:2011ip,Barenboim:2016shh}. These limits apply at  later times, when $T< 1$ MeV, whereas the production of sterile neutrinos in our analysis effectively ceases   the time at which  $T \simeq 10$ MeV. Therefore,  since resonant production partially depletes the lepton number,  the initial lepton number can be significantly larger than the  mentioned constraints.  

At $T< 10$ MeV,  active neutrino oscillations become efficient and could equilibrate lepton asymmetries across flavors before the $T\simeq$ 1 MeV~\cite{Dolgov:2002ab}.  This equilibration is not guaranteed, and if it happens could convert excess lepton asymmetry into entropy injection reheating the primordial plasma and causing entropy dilution (see e.g. ~\cite{Lattanzi:2024hnq, Froustey:2024mgf, Domcke:2025lzg}). Thus, active neutrino flavor oscillations can complicate the relation between primordial lepton asymmetry at $T> 10$ MeV, and its value at the BBN-era and later. For our analysis, we evolve lepton asymmetries down to a reference temperature of $3$ MeV, approximately when the active neutrinos decouple, and omit flavor conversions among active neutrinos, as they have negligible impact on the sterile neutrino population already frozen-in by $T\simeq 10$ MeV.
 
We assume that the sterile neutrino number density $n_{\nu_s}$ is initially negligible after inflationary reheating
and remains small  with respect to the thermal population of active neutrinos, $n_{\nu_s} \ll n_{\nu_\alpha}$ throughout the evolution.  
Thus, sterile neutrinos  never reach equilibrium, and $L_\alpha$,  defined  in Eq.~\eqref{eq:leptonnum},  is continuously evolving  during resonant production. 
 
Sterile neutrino resonant production can occur either through coherent MSW-like oscillations (Shi-Fuller mechanism~\cite{Shi:1998km}), or through collisional decoherence with in-medium maximal mixing ($\sin^2 2\theta_m= 1$). In this work, we focus on the regime where collisional decoherence dominates, and the Boltzmann equation directly applies. This regime corresponds to most of the parameter space of interest for sterile neutrino masses $m_s>\textrm{keV}$, where the conditions for Shi–Fuller production—coherence and adiabaticity—are not satisfied during the resonance~\cite{Abazajian:2001nj,Gelmini:2020duq} (see figures in Ref.~\cite{Gelmini:2019clw} for parameter space of Shi-Fuller production). 

In the parameter space we consider, sterile neutrino production is therefore well described by the Boltzmann Eq.~\eqref{eq:boltzmannp}, with resonant enhancement incorporated via the in-medium mixing angle in Eq.~\eqref{eq:sin2thetam}.

\subsection{Lepton asymmetry evolution}
\label{sec:lepasym}

The evolution of the lepton asymmetry can play a key role in controlling sterile neutrino production in the early Universe. It determines the transition from resonant to non-resonant active-sterile oscillations, shaping the sterile neutrino momentum distribution. Here, we consider a streamlined treatment of lepton asymmetry evolution, using generalized variables $T_g$ to simplify the calculations. We account for antineutrino contributions, which are negligible during resonance but can become important in the non-resonant regime. This approach enables accurate tracking of lepton asymmetry depletion and its impact on the production of distinct sterile neutrino populations.

The rate of change  of the lepton number \ in terms of the conversion rate $\Gamma$ in Eq.~\eqref{eq:totalrate}  is (given as $dL/dt$ in Ref.~\cite{PhysRevD.59.113001})
\begin{align}
\label{eq:dLdt}
    \frac{dL_\alpha}{dT_g} \simeq&~ \frac{g_*(T_g)}{2n_\gamma(T_g)}\int d\epsilon_g ~ \frac{\epsilon_g^2}{HT_g}
    \left\{\Gamma(\nu_\alpha\rightarrow \nu_s; \epsilon_g,T_g)f_{\nu_\alpha}(\epsilon_g,T_g)
    -\Gamma(\bar{\nu}_\alpha\rightarrow \bar{\nu}_s;  \epsilon_g,T_g)f_{\bar{\nu}_\alpha}(\epsilon_g,T_g)\right. \notag\\
    &\left. -\Gamma(\nu_s\rightarrow \nu_\alpha; \epsilon_g,t)f_{\nu_s}(\epsilon_g,T_g)
    +\Gamma(\bar{\nu}_s\rightarrow \bar{\nu}_\alpha; \epsilon_g,T_g)f_{\bar{\nu}_s}(\epsilon_g,T_g)\right\}~.  
\end{align}
Since the conversion probability is the same for the forward and reverse processes $P(\nu_\alpha\rightarrow \nu_s)=P(\nu_s\rightarrow\nu_\alpha)$, 
it follows that $\Gamma(\nu_\alpha\rightarrow \nu_s)=\Gamma(\nu_s\rightarrow \nu_\alpha)$, and similarly $\Gamma(\bar{\nu}_\alpha\rightarrow \bar{\nu}_s)=\Gamma(\bar{\nu}_s\rightarrow \bar{\nu}_\alpha)$ for their antineutrino counterparts. Eq.~\eqref{eq:dLdt} can then be simplified as
\begin{equation}
\label{eq:dLdtv2}
\begin{aligned}
    \frac{dL_\alpha}{dT_g}
  \simeq&~ \frac{g_*(T_g)}{2n_\gamma(T_g)}\int d\epsilon_g   ~ \frac{\epsilon_g^2}{HT_g}
    \left\{\Gamma(\nu_\alpha\rightarrow \nu_s; \epsilon_g,T_g)[f_{{\nu}_\alpha}(\epsilon_g,T_g)-f_{{\nu}_s}(\epsilon_g,T_g)]\right.\\
   &\left. -\Gamma(\bar{\nu}_\alpha\rightarrow \bar{\nu}_s; \epsilon_g,T_g)
   [f_{\bar{\nu}_\alpha}(\epsilon_g,T_g)-f_{\bar{\nu}_s}(\epsilon_g,T_g)]\right\}~.
   \end{aligned}
\end{equation}

\begin{figure}[t]
    \centering  \includegraphics[width=0.49\linewidth]{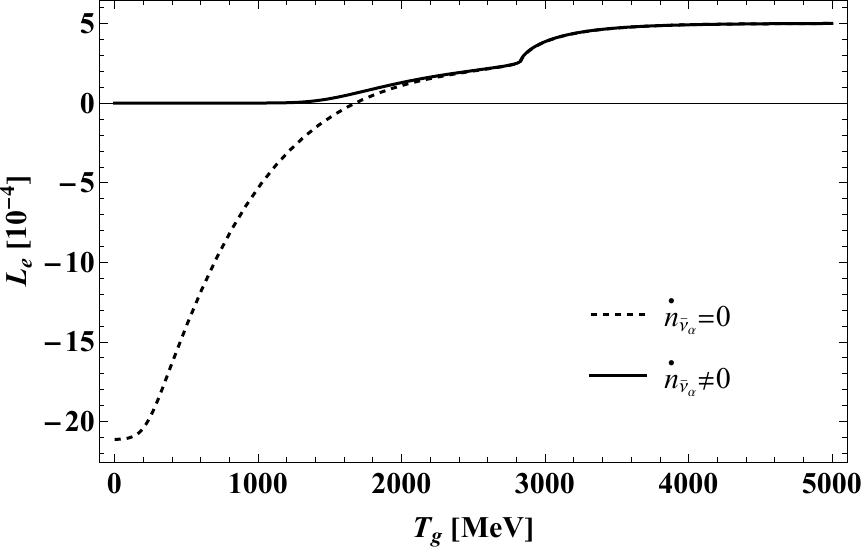}
    \caption{Lepton asymmetry as a function of $T_g$, calculated with (solid line)  
    and without (dashed line) the antineutrino contribution for $m_s=10$ keV, $\sin^22\theta=2\times10^{-9}$, and initial lepton number $L_{e,0}=0.5\times 10^{-3}$.}
    \label{fig:antinu-L}
\end{figure}
 
A distinction of our work compared to some previous studies~(e.g.~\cite{Kishimoto_2008,Abazajian:2014gza,Kishimoto:2006zk}),  
is the explicit inclusion of antineutrino conversions in the evolution of the lepton number $\Dot{L}_\alpha$ (antineutrino conversions have also been included in several previous studies, e.g.~\cite{Horner:2023cmc,Kasai:2024diy}).
While the antineutrino conversion rate is suppressed 
when the lepton asymmetry is large and positive, making the antineutrino contribution negligible during resonant production of sterile neutrinos, their role becomes significant in the non-resonant regime. In Fig.~\ref{fig:antinu-L} we display an example of the lepton number as function of $T_g$, calculated with (solid line)  and without (dashed line) the antineutrino contribution. As shown in the figure, antineutrino conversions can substantially reduce the rate of lepton number depletion after the resonant production phase ends (indicated in the figure by the separation of the dashed and solid lines). During  
resonant conversion, at high energies in the figure, $\bar{\nu}_\alpha$ transitions are strongly suppressed, resulting in the nearly identical evolution of $L_{\alpha}$ whether or not antineutrino conversions are included. 
At lower temperatures, after resonance has ended and much of the initial lepton asymmetry has been depleted, the neutrino and antineutrino conversion rates become comparable in magnitude but opposite in sign. This drives the lepton number toward zero, as evident in the low-temperature behavior shown in Fig.~\ref{fig:antinu-L}.

By implementing appropriately small adaptive time steps in our numerical calculations, we ensure accurate tracking of the antineutrino contribution to the lepton asymmetry evolution. This prevents the net driving lepton number $\mathcal{L_\alpha}$ from becoming negative. In contrast, some previous studies ~\cite{Kishimoto_2008,Horner:2023cmc} assumed non-zero lepton asymmetries for other, non-interacting flavors. This enabled the lepton asymmetry of the interacting flavor, here $\nu_e$, to fall below zero while maintaining a positive total $\mathcal{L_\alpha}$. In our work, since we assume $L_{\mu,\tau}=0$ initially, the electron lepton asymmetry remains positive $L_e>0$ throughout the Boltzmann evolution.

\begin{figure}[t]
    \centering
    \includegraphics[width=0.49\linewidth]{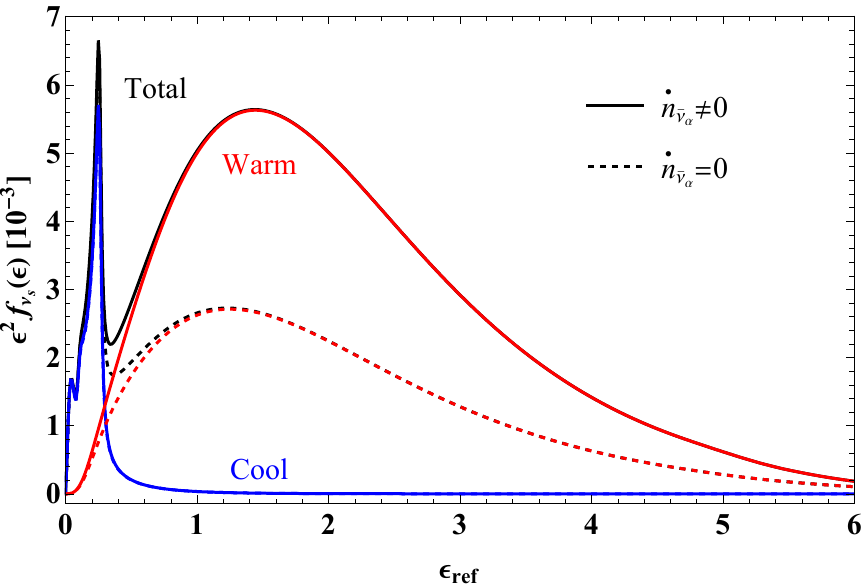}
    \caption{Comparison of momentum distributions $\epsilon^2f_{\nu_s}(\epsilon)$ at $T=3$ MeV  calculated with (solid) or without (dashed) the antineutrino contribution for  $m_s=10$ keV, $\sin^22\theta=2 \times10^{-9}$, and $L_{e,0}=5\times 10^{-4}$, with reference $\epsilon(t)=\epsilon_{\rm ref}$ at the time.  
    Blue and red curves indicate the cool and warm sterile neutrino populations, respectively, as discussed in Sec.~\ref{ssec:twopop}, while the black curves show the total momentum distribution. The inclusion of $\bar{\nu}_\alpha\xrightarrow{}\bar{\nu}_s$ conversions effectively doubles the (non-resonant) warm production,
     slightly  the resonant cool production through the injection of positive lepton asymmetry from antineutrino transitions.
    }
    \label{fig:antinu-fs}
\end{figure}

Fig.~\ref{fig:antinu-fs} illustrates the impact of including antineutrino conversions on the sterile neutrino momentum distribution. The figure shows the sterile neutrino momentum distribution  at $T= 3$ MeV as function of $\epsilon(t)= \epsilon_{\rm ref}$ at the time.  In Sec.~\ref{sec:separatefs} we
define a cooler sterile neutrino population (blue solid line in Fig.~\ref{fig:antinu-fs}) produced earlier through resonant oscillations and a warmer population (red solid line in Fig.~\ref{fig:antinu-fs}) produced later through non-resonant oscillations (the total distribution is the sum of both, shown with the black solid line). The warm population
is approximately doubled due to the inclusion of antineutrino conversions (the result of not including them is shown with the red dashed line), since the Boltzmann equation is nearly symmetric between $\nu_\alpha$ and $\bar{\nu}_\alpha$ in this regime. 
The $\bar{\nu}_\alpha\xrightarrow{}\bar{\nu}_s$ conversions also act to slow the depletion of the lepton asymmetry, effectively maintaining a higher asymmetry for longer. For resonant (and non-resonant) cool population production, defined in Sec.~\ref{ssec:twopop}, the inclusion of antineutrino conversions leads to a modest enhancement of the sterile neutrino distribution, with a peak amplitude increase of about $\mathcal{O}(10)\%$. This arises from the injection of positive lepton asymmetry due to $\bar{\nu}_\alpha\xrightarrow{}\bar{\nu}_s$ transitions, which temporarily reduces the rate of lepton number depletion and slightly boosts production near the resonance.

\subsection{Production regimes}
\label{sec:separatefs}

We systematically analyze sterile neutrino production in the presence of a lepton asymmetry, explicitly distinguishing between different momentum components. Unlike some previous works that associate the cool sterile neutrino population solely with resonant production. Besides the typical cool resonantly produced population, we identify an additional non-resonant cool population alongside a non-resonant warm component. Thus, by tracking production across momentum and temperature, we define three distinct populations—resonant cool, non-resonant cool, and warm—and determine the precise conditions under which each is generated. This approach provides a comprehensive and systematic framework for understanding mixed sterile neutrino DM scenarios.
\begin{figure}
    \centering
\includegraphics[width=0.49\linewidth]{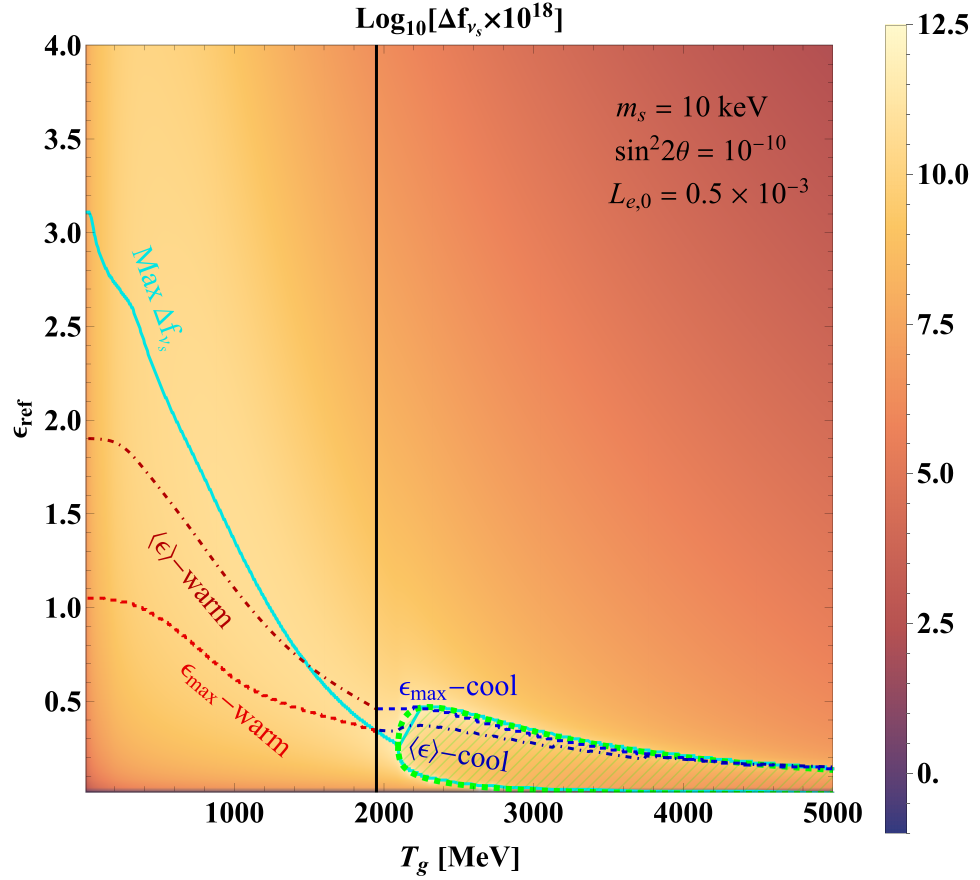}  \includegraphics[width=0.49\linewidth]{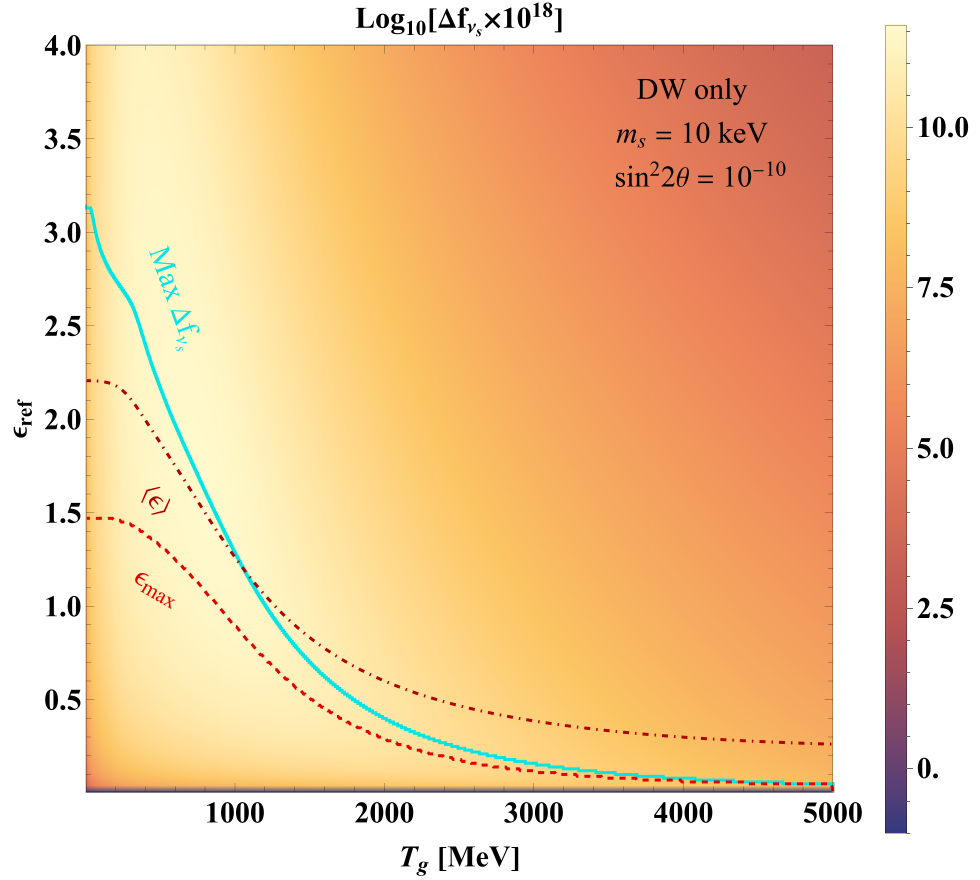}
    \caption{Color-coded density plot of $\Delta f_{\nu_s}(\epsilon_{\rm ref}, T_g)$, defined in Eq.\eqref{eq:Delf}, redshifted to the reference temperature $T = 3$ MeV, as a function of $T_g$ and $\epsilon_{\rm ref}$ (i.e., $\epsilon(t)$ evaluated at $T=3$ MeV), for $m_s = 10$ keV and $\sin^2 2\theta = 10^{-10}$. The solid green line indicates the value of $\epsilon_{\rm ref}$ where $\Delta f_{\nu_s}(\epsilon_{\rm ref}, T_g)$ reaches its maximum at each $T_g$. The evolution of the average momentum $\langle \epsilon_{\rm ref} \rangle$ (dot-dashed lines) and the momentum at which the distribution $f_{\nu_s}$ is maximal, $\epsilon_{\rm max}$ (dashed lines), are shown separately for the cool and warm populations, as defined in Sec.~\ref{ssec:twopop}. [Left] Production with an initial lepton asymmetry $L_{e,0} = 0.5 \times 10^{-3}$, generating a cool population (blue) with $\langle \epsilon_{\rm ref} \rangle \simeq 0.34$ and a warm population (red) with $\langle \epsilon_{\rm ref} \rangle \simeq 1.90$. The separation between the two populations occurs at $T_g \simeq 1950$ MeV, where the green and dot-dashed lines cross (see Sec.~\ref{ssec:twopop}). The two resonance branches (dashed green) and the region where $F_{\rm res} < 0$ (hatched green), corresponding to potential non-resonant cool production, are also shown. The dominant production occurs near the resonance, resulting in a primarily cool sterile neutrino population. [Right] The $L_{e,0} = 0$ case (DW production) is shown for comparison. In this case, most production happens around $T_g \simeq 500$ MeV, yielding a warm sterile neutrino population with $\langle \epsilon_{\rm ref} \rangle \simeq 2.28$.
    }
    \label{fig:fsdens-10}
\end{figure}
 
\begin{figure}
\centering
\includegraphics[width=0.465\linewidth]{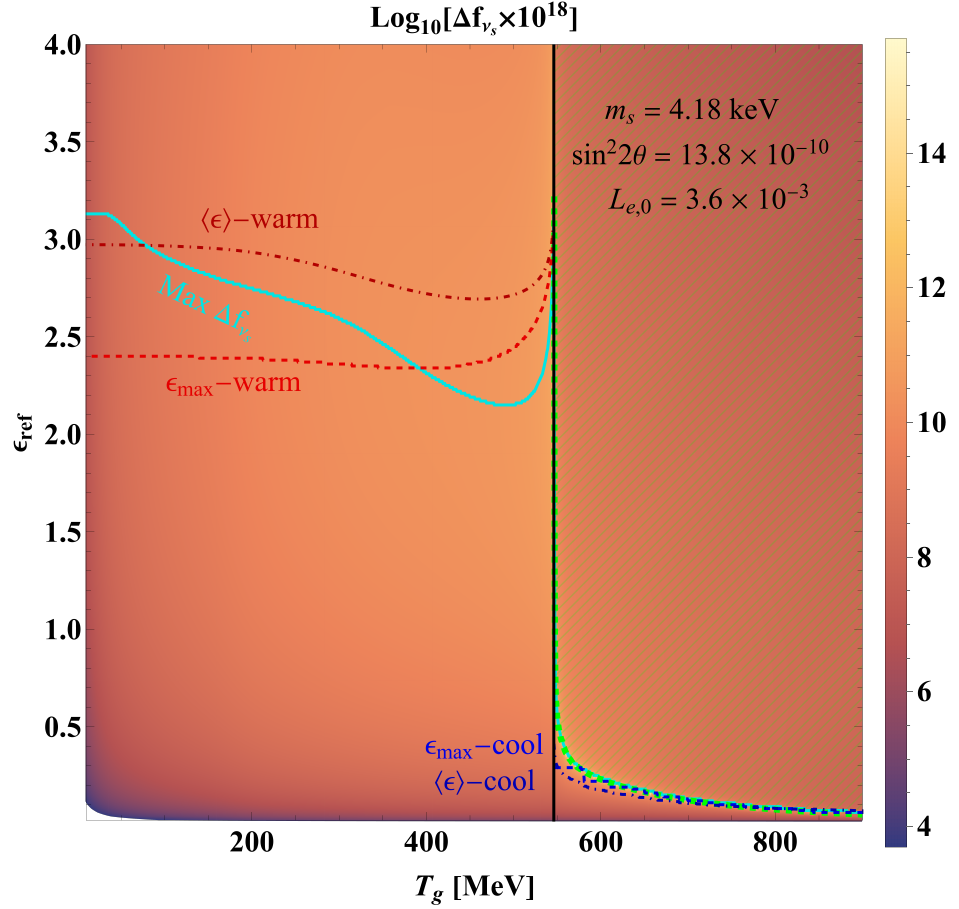}
\includegraphics[width=0.49\linewidth]{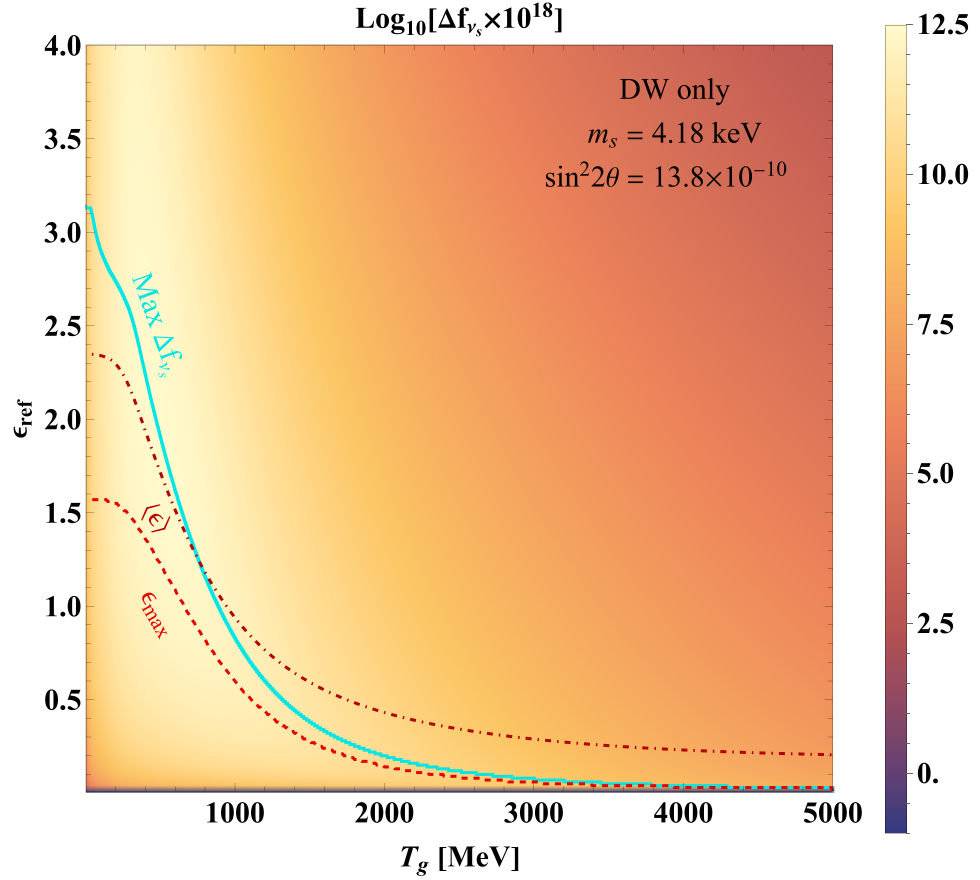}
\caption{Same as Fig.~\ref{fig:fsdens-10}, but for $m_s = 4.18$ keV, $\sin^2 2\theta = 13.8 \times 10^{-10}$, and $L_0 = 3.65 \times 10^{-3}$. In the left panel, resonant cool production is suppressed, and non-resonant cool production dominates, in contrast to the $m_s = 10$ keV case shown in Fig.~\ref{fig:fsdens-10}. This highlights the parameter dependence of the sterile neutrino production mechanism.}
\label{fig:fsdens-4-2}
\end{figure}

In the presence of a significant initial lepton asymmetry, the condition for resonant production, $F_{\rm res} = 0$ (see Eq.~\eqref{eq:Fres}), can be satisfied. The condition $F_{\rm res} = 0$ is quadratic in $\epsilon_g$, so there are two momentum values, $\epsilon_{g,\rm res}^-$ and $\epsilon_{g,\rm res}^+$, undergoing resonance at any one time (or temperature). These two solutions merge at a temperature $T_{\rm end}$, i.e. at the root of the discriminant, below which the resonance condition is no longer satisfied. 

Generically, resonant production occurs earlier at higher temperatures, while subsequent non-resonant production can occur once the lepton asymmetry is sufficiently depleted.
The type of production depends on the dominant term in the denominator of the effective mixing angle $\sin^2 2\theta_m$ of Eq.~\eqref{eq:sin2thetam}. At higher temperatures, the thermal potential $T_g V_T\propto -T_g^6$ dominates and suppresses production. For sizable lepton asymmetry, $\mathcal{L_\alpha}\gtrsim 10^{-5}$, the density potential $T_g V_D\propto \mathcal{L}_\alpha T_g^4$ can cancel $V_T$ at  $T \gtrsim\mathcal{O}($GeV)  
for sterile neutrino masses of $\mathcal{O}($keV). Since the sign of the density potential depends on whether neutrinos or anti-neutrinos are considered, only one of them will experience enhancement in production due to resonance. This asymmetric production leads to depletion of lepton asymmetry via Eq.~\eqref{eq:dLdt},  which eventually becomes too small to induce resonance. At lower temperatures $T\simeq \mathcal{O}(100~\textrm{MeV})$, non-resonant production may occur unimpeded if the lepton asymmetry is sufficiently decreased such that $T_gV_D/m_s^2 \ll 1$. In Sec.~\ref{ssec:twopop}, we discuss in more detail the conditions for non-resonant production to 
happen after resonant production, forming two populations of sterile neutrinos with very different characteristic momenta.

In our numerical calculation of sterile neutrino production (detailed in App.~\ref{sec:num-cal}), we integrate Eq.\eqref{eq:boltz-eg-approx} over a small temperature interval from $T_g - \delta T_g$ to $T_g$ to obtain the change in the sterile neutrino distribution function $f_{\nu_s}$ during that step (see Eq.~\eqref{deltaf-appB} for details) 
\begin{eqnarray}
\label{eq:Delf}
    \Delta f_{\nu_s}&=&-\int_{T_{g}-\delta T_{g}}^{T_g} \frac{dT_g'}{H(T_g')T_g'} \Gamma(\nu_\alpha\rightarrow\nu_s;\epsilon_g,T_g')f_{\nu_\alpha}(\epsilon_g,T_g')~.
\end{eqnarray}
The final momentum distribution is found by summing all these contributions (see Eq.~\eqref{f-as-sun-of-deltaf-appB}).  

Fig.~\ref{fig:fsdens-10} illustrates the evolution of $\Delta f_{\nu_s}$, with its value redshifted to $T = 3$ MeV and displayed in the $(T_g, \epsilon_{\rm ref})$ parameter space with a color code (lighter tones correspond to larger values). Here, $\epsilon_{\rm ref}$ denotes the scaled momentum $\epsilon(t)$ evaluated at $T = 3$ MeV. The left panel shows the case of a large initial lepton asymmetry, while the right panel corresponds to negligible asymmetry, representing pure DW production. We indicate the maximum of $\Delta f_{\nu_s}$ by a green line labeled ``Max $\Delta f_{\nu_s}$".

\begin{figure}[t]
    \centering
    \includegraphics[width=0.49
    \linewidth]{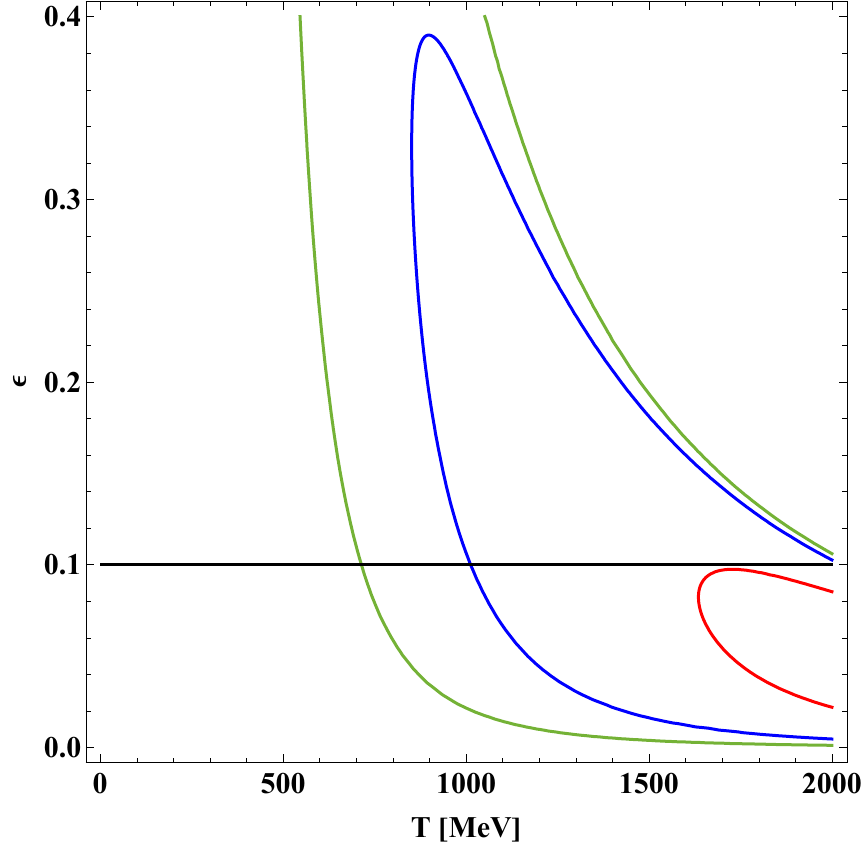}
    \caption{Temperature evolution of the two resonance branches, $\epsilon_{\rm res}^{\pm}$, corresponding to the solutions of $F_{\rm res} = 0$, for an approximately constant lepton asymmetry $\mathcal{L} \simeq \mathcal{L}_0 = 10^{-3}$. Results are shown for $m_s = 10$ keV (green), $m_s = 20$ keV (blue), and $m_s = 40$ keV (red). Here, $\epsilon$ is evaluated at the time of production, not redshifted to the reference temperature, since this figure illustrates the production condition. This plot also demonstrates the condition for resonant production to occur, $m_s < m_{\rm reslim}$. For a given momentum, resonance production is possible if $\epsilon_{\rm res}^{\pm}$ intersects the constant $\epsilon$ line, shown in black for $\epsilon = 0.1$.
    }
    \label{fig:msres}
\end{figure}

In the left panel of Fig.~\ref{fig:fsdens-10}, as the temperature and lepton asymmetry decrease, the two local maxima of $\Delta f_{\nu_s}$ evolve along curves that track the two resonance branches of the condition $F_{\rm res} = 0$. Due to the presence of the damping term in the denominator of Eq.~\eqref{eq:sin2thetam}, the maximum of $\Delta f_{\nu_s}$ occurs at $\epsilon_{\rm ref}$ values slightly smaller than the exact resonance momentum.
Temperature behavior of 
resonance branches is illustrated in~Fig.~\ref{fig:msres}.
After the two resonance branches merge, the condition $F_{\rm res} = 0$ is no longer satisfied, marking the end of resonant production. However, the remaining lepton asymmetry still modifies the production rate, influencing non-resonant contributions. We continue the green line in the left panel of Fig.~\ref{fig:fsdens-10} to lower temperatures, showing the $\epsilon_{\rm ref}$ corresponding to the maximum of $\Delta f_{\nu_s}$. This continuation closely follows the region where $F_{\rm res} > 0$ is minimized.

\begin{figure}
    \centering
\includegraphics[width=0.49\linewidth]{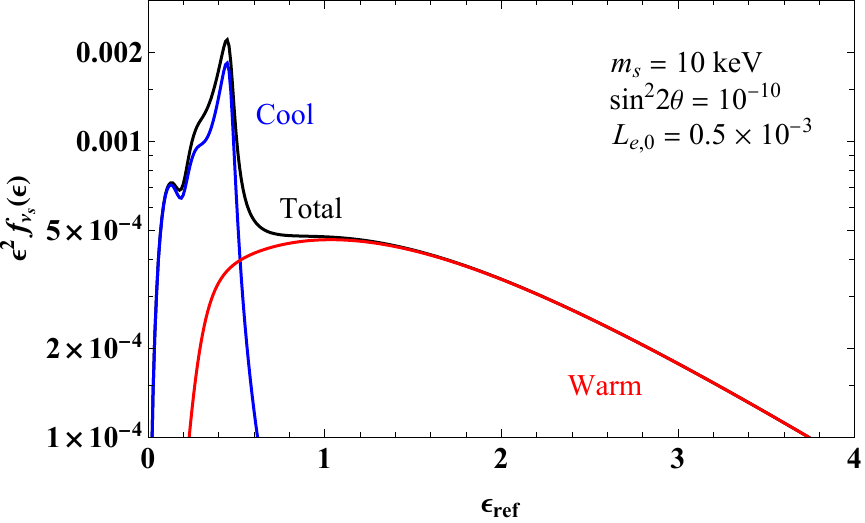}    \includegraphics[width=0.49\linewidth]{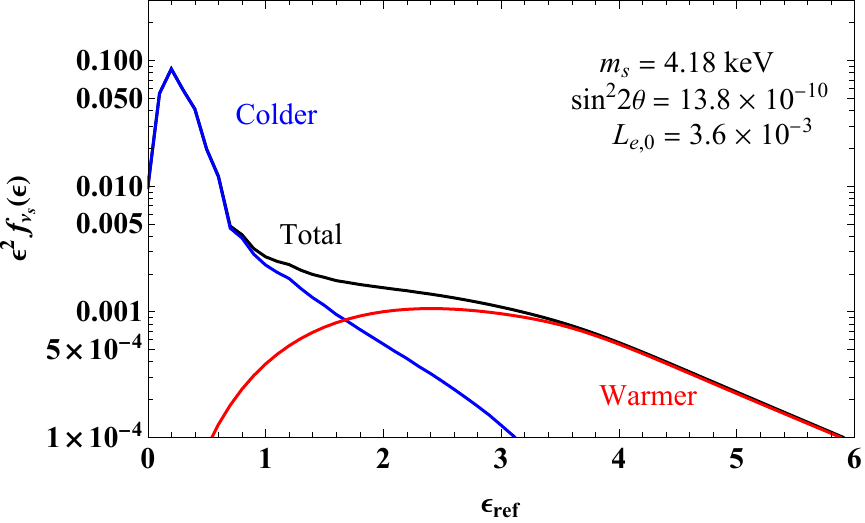}
    \caption{Sterile neutrino momentum density, $\epsilon^2 f_{\nu_s}(\epsilon)$, for the cool (blue) and warm (red) populations defined in Sec.~\ref{ssec:twopop}, along with the total population (black).~[Left] Model with $m_s = 10$ keV, $\sin^2 2\theta = 10^{-10}$, and initial lepton asymmetry $L_{e,0} = 0.5 \times 10^{-3}$. The cool population exhibits three peaks, the two outer peaks at $\epsilon_{\rm ref} \simeq 0.1$ and $\epsilon_{\rm ref} \simeq 0.5$ originate from the two resonance branches $\epsilon_{\rm res}^{\pm}$, while the central peak at $\epsilon_{\rm ref} \simeq 0.26$ results from a brief period of non-resonant cool production immediately after the end of resonant production, where $F_{\rm res} \gtrsim 0$.~[Right] Model with $m_s = 4.18$ keV, $\sin^2 2\theta = 13.8 \times 10^{-10}$, and $L_{e,0} = 3.65 \times 10^{-3}$. In this case, the cool population spectrum has only a single peak, corresponding to non-resonant cool production. Unlike the left panel, no resonant peaks are present. The results are binned in intervals of $\Delta \epsilon_{\rm ref} = 0.1$ to smooth out numerical fluctuations due to the rapid evolution of the resonance curves, which requires extremely small temperature steps.
}
    \label{fig:fscomp}
\end{figure}

The two resonance branches partition the $(T_g, \epsilon_{\rm ref})$ parameter space into three distinct regions, corresponding to the production of three separate sterile neutrino populations that include resonantly produced cool component, non-resonantly produced cool component and   non-resonantly produced warm component:
\begin{enumerate}
    \item {\it Resonant production of cool population}\\ 
    The resonance effect significantly enhances sterile neutrino production along the two resonance branches (green lines in the left panel of Fig.~\ref{fig:fsdens-10}), resulting in two distinct peaks in the sterile neutrino momentum distribution (corresponding to the two outer peaks in the left panel of Fig.~\ref{fig:fscomp}). The average momentum of this resonantly produced population is typically $\langle\epsilon_{\rm ref}\rangle \lesssim 1$, substantially cooler than that of the non-resonant  DW  population. For sterile neutrino masses of $m_s \simeq \mathcal{O}(\textrm{keV})$, significant resonant production occurs for initial lepton asymmetries $L_{e,0} \gtrsim 10^{-5}$. The left panel of Fig.~\ref{fig:fsdens-10} illustrates this case. As the temperature decreases, the two resonance branches eventually merge at a temperature we denote $T_{\rm end}$, marking the termination of resonant production.

    \item  {\it Non-resonant production of cool population}\\ 
    Between the two resonance branches, the density potential $V_D$ dominates over the other terms in $F_{\rm res}$, leading to $F_{\rm res} < 0$. Although the resonance condition is not satisfied in this region, the magnitude of $F_{\rm res}$ can still reduced compared to the case with zero lepton asymmetry ($\mathcal{L} = 0$). This results in moderately enhanced production of a non-resonant cool population, characterized by an average momentum $\langle\epsilon_g\rangle \lesssim 1$. For small sterile neutrino masses, $m_s \lesssim \textrm{keV}$, the two resonance branches are widely separated, with $\epsilon_{g,\rm res}^- \rightarrow 0$ and $\epsilon_{g,\rm res}^+ \gtrsim 5$. Due to the suppressed phase-space occupancy of active neutrinos at these extremal momenta, production along the resonance curves is diminished. Consequently, for small masses, the production between the resonance branches can become the dominant contribution, as shown in the left panel of Fig.~\ref{fig:fsdens-4-2}. In this case, the green line tracing the maximum of $\Delta f_{\nu_s}$ corresponds to the peak of the production rate of this non-resonant cool component.  
    Additionally, some non-resonant cool production can occur in the region directly to the left of the merged resonance curves, at $T \lesssim T_{\rm end}$. Although the resonance condition is not satisfied here ($F_{\rm res} \gtrsim 0$), the remaining lepton asymmetry still enhances sterile neutrino production, leading to the formation of a third peak between the two resonance peaks in the distribution, as shown in the left panel of Fig.~\ref{fig:fscomp}.
    \item {\it Non-resonant production of warm population}\\ 
 Above and to the left of the resonance branches in the left panel of Fig.~\ref{fig:fsdens-10}, the function $F_{\rm res}$ is dominated by either the $\cos2\theta$ term or the thermal potential $V_T$. In this region of $(\epsilon_g, T_g)$, sterile neutrino production proceeds non-resonantly. As the lepton asymmetry decreases, this production approaches the standard DW scenario. As discussed in Sec.~\ref{ssec:twopop}, significant DW production can occur after resonant production if $T_{\rm max} < T_{\rm end}$.  
 For small initial lepton asymmetries, $L_{e,0} < 10^{-5}$, the resonance region shrinks to a narrow corner of parameter space, and non-resonant production dominates. 
 In this case, the maximum of $\Delta f_{\rm \nu_s}$ approaches $\epsilon_{\rm ref}\simeq3.13$ before the production finishes. 
 We demonstrate this in the right panels of Figs.~\ref{fig:fsdens-10} and~\ref{fig:fsdens-4-2}, where the lepton asymmetry is negligible, and the sterile neutrino production is entirely non-resonant.
Similar behavior is observed in the left panels of these figures at temperatures $T < T_{\rm end}$, where the lepton asymmetry has been depleted and non-resonant production proceeds unimpeded.
\end{enumerate}
The sterile neutrino mass $m_s$, mixing angle $\sin^2 2\theta$, and initial lepton asymmetry $\mathcal{L}_\alpha$ together determine which of the populations—resonant cool, non-resonant cool, or warm—dominates the final momentum distribution.
 
 \subsection{Two populations: cool and warm}
\label{ssec:twopop}

\begin{figure}
    \centering
    \includegraphics[width=0.49
    \linewidth]{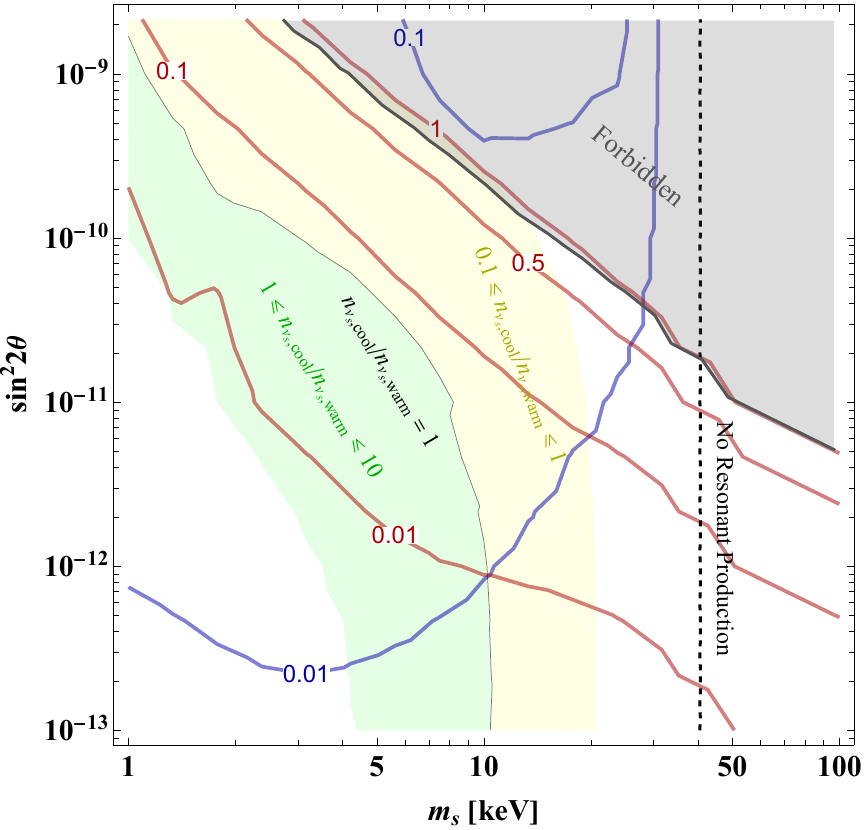}
    \caption{Contours of equal DM density fraction in sterile neutrinos, shown separately for the cool (blue) and warm (red) populations, in the $(m_s,~\sin^2 2\theta)$ plane for $\mathcal{L}_0 = 10^{-3}$. The value of $f_{\rm s,DM}$ for each population is indicated along the corresponding contour lines.
The dashed black vertical line marks the upper mass limit for significant resonant production, $m^{\rm mod}_{\rm reslim}$ from Eq.~\eqref{eq:msupper}, to the right of this line resonant production no longer occurs. The solid black slanted line indicates where sterile neutrinos account for all of the dark matter, $f{\rm s,DM}^{\rm cool} + f_{\rm s,DM}^{\rm warm} \simeq 1$. The gray region above this line is excluded since $f_{\rm s,DM} > 1$.
Colored bands indicate the region where both warm and cool populations are sizable, defined by $0.1 \lesssim n_{\nu_s, \rm cool}/n_{\nu_s, \rm warm} \lesssim 10$. In the green band, the cool component dominates, in the yellow band, the warm component dominates, and along the thin black line separating them, the two populations contribute equally.
These mixed-population regions are largely contained between the analytic bounds $m^{\rm mod}_{\rm non-res} = 0.29$ keV (outside the plotted range) and $m^{\rm mod}_{\rm reslim} = 40.6$ keV, as defined by Eqs.~\eqref{eq:mslower} and~\eqref{eq:msupper}, consistent with the range proposed in~Ref.~\cite{Gelmini:2019clw}.
    }
    \label{fig:fres-fnon-1}
\end{figure}

\begin{figure}
    \centering
    \includegraphics[width=0.49
    \linewidth]{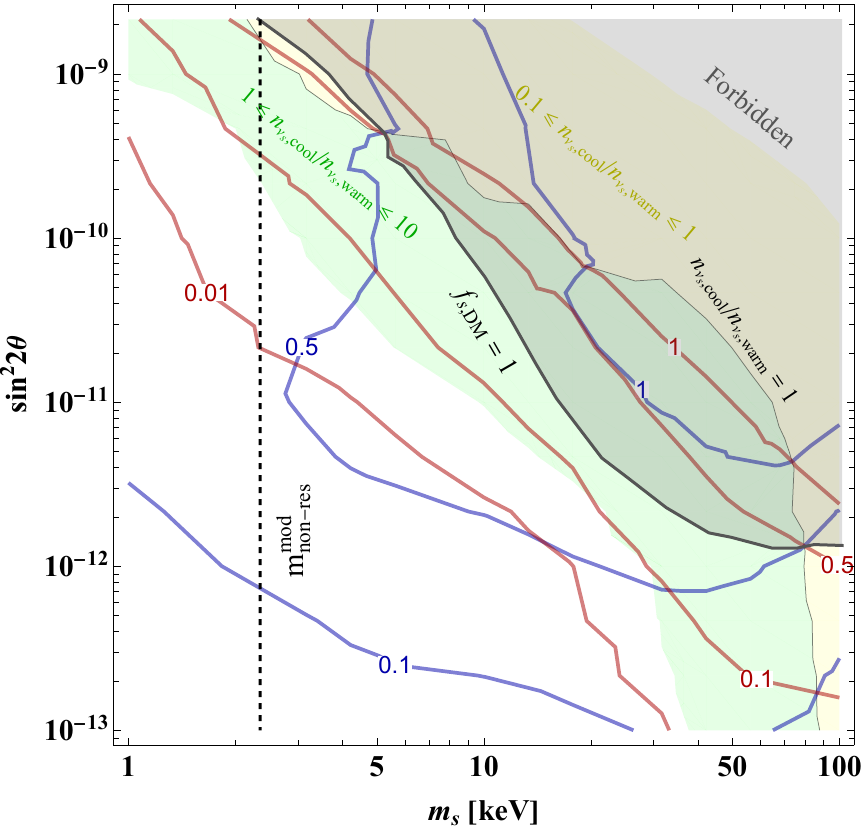}
    \caption{Same as Fig.~\ref{fig:fres-fnon-1}, but for $\mathcal{L}_0 = 4 \times 10^{-3}$. The colored bands indicate the region where both cool (green) and warm (yellow) sterile neutrino populations are sizable, defined by $0.1 \lesssim n_{\nu_s, \rm cool}/n_{\nu_s, \rm warm} \lesssim 10$. The thin black line separating the bands corresponds to equal contributions from the two populations. The mixed-population region largely lies between $m^{\rm mod}_{\rm non-res} = 2.35$ keV (shown by the vertical black dashed line) and $m^{\rm mod}_{\rm reslim} = 320$ keV (that is outside the plot range), consistent with the analytic range defined by Eqs.~\eqref{eq:mslower} and~\eqref{eq:msupper}, and similar to that suggested in~Ref.~\cite{Gelmini:2019clw}.
    }
\label{fig:fres-fnon-2}
\end{figure}

We explicitly and systematically explore the conditions under which sterile neutrino production results in two sizable populations: one cool and one warm. Previously in our earlier work~\cite{Gelmini:2020ekg} we suggested approximate criteria for the coexistence of resonant and non-resonant populations. Here, we refine and extend these conditions both analytically and numerically. We confirm with numerical analysis (see Figs.~\ref{fig:fres-fnon-1} and \ref{fig:fres-fnon-2}) that the two-population regimes are indeed contained within the analytic ranges we derive. This framework provides a systematic and quantitative mapping of the parameter space relevant for mixed cold and warm sterile neutrino DM scenarios.

As discussed in Sec.~\ref{sec:separatefs}, the resonantly produced population is typically cooler  than the non-resonantly produced population. 
In the left panel of  Fig.~\ref{fig:fsdens-10} for example, immediately after resonant production ends, the rate of the non-resonant production is maximum in between the two resonant peaks. We denote this as ``non-resonant cool" production. After the lepton asymmetry is further diminished, the non-resonant production peak shifts toward higher momenta, ending at 
$\epsilon_{\rm ref}\simeq 3.13$.  

In this work, we focus on the production of multiple sizable populations of the same sterile neutrino species, including one colder population produced earlier during resonance, and one warmer population produced later after resonant production has ceased. We separate the two populations at the temperature where the value of $\epsilon_{\rm ref}$ corresponding to the maximum of $\Delta f_{\nu_s}$ (indicated by the green line labeled ``Max $\Delta f_{\nu_s}$" in Fig.~\ref{fig:fsdens-10}) after resonance production ceases (when $T < T_{\rm end}$) becomes larger than the maximum average momentum $\langle \epsilon_{\rm ref}\rangle$ during the resonant production phase (shown by the blue dot-dashed line in Fig.~\ref{fig:fsdens-10}).  This crossing point defines the separation between the cool and warm components and occurs at $T_g \simeq 1950$ MeV in the left panel of Fig.~\ref{fig:fsdens-10}. This approach guarantees that sterile neutrinos produced at later times are systematically warmer than those produced earlier. The temperature evolution of the average momentum $\langle \epsilon_{\rm ref}\rangle$ is shown separately for the cool and warm populations in Figs.~\ref{fig:fsdens-10} and~\ref{fig:fsdens-4-2}, with the cool component indicated by the blue dot-dashed line and the warm component by the red dot-dashed line. Additionally, the values of $\epsilon_{\rm max}$, corresponding to the peak of the density spectrum $f_{\nu_s}$ for each population, are shown as blue and red dashed lines for the cool and warm components, respectively.

In general, for sterile neutrino masses in the keV range, both colder and warmer populations are produced. However, depending on the model parameters, either component can become subdominant or negligible. For each value of the initial lepton asymmetry $\mathcal{L}_0$, the $(m_s, \sin^2 2\theta)$ parameter space naturally divides into three distinct regions: one where the production is dominated by the cool component (cold-only), one where both populations contribute significantly (two-population), and one where the warm component dominates (warm-only).

In Ref.~\cite{Gelmini:2019clw}, an approximate analytic condition was proposed to identify the parameter space where two sizable sterile neutrino populations can form. It was argued that, since the DW production rate has a sharp maximum at $T_{\rm max}$ (see Eq.~\eqref{eq:Tmax}), the production of a significant warm population after resonant production is possible if $T_{\rm max} < T_{\rm res}$, where $T_{\rm res}$ is the temperature corresponding to the lower resonance branch $\epsilon_{\rm g,res}^-$. This condition ensures that non-resonant production remains active after the depletion of the lepton asymmetry. Using the relation $T_{\rm res} \simeq 18.8~\mathrm{MeV} (\epsilon(t)\mathcal{L}_0)^{-1/4}(m_s/\mathrm{keV})^{1/2}$, this criterion translates into a lower limit on the sterile neutrino mass, $m_s \gtrsim m_{\rm non-res} \simeq 3.59 \times 10^4~\mathrm{keV}  \epsilon(t)^{2/3} \mathcal{L}_0^{2/3}$.
In Ref.~\cite{Gelmini:2019clw}, this was combined with the upper mass limit for resonant production, defined by the condition that $F_{\rm res}=0$ has a solution. For fixed $\epsilon=1$, this upper bound is $m_s < m_{\rm reslim} \simeq 4.03 \times 10^5\mathrm{keV}  \mathcal{L}_0^{3/2}$.
Together, these bounds define the mass range where a two-population scenario is possible: $m_{\rm non-res} \lesssim m_s < m_{\rm reslim}$.
With our present numerical calculations, we confirm that significant cool and warm sterile neutrino populations are produced largely within this analytically predicted range.

We improve upon the approximate analytic conditions of Ref.~\cite{Gelmini:2019clw} by employing $T_{\rm end}$ in place of $T_{\rm res}$, that is by requiring $T_{\rm max} < T_{\rm end}$. This leads to a new modified estimate of the lower mass bound, denoted $m^{\rm mod}_{\rm non-res}$, and also allows us to compute a refined upper limit for resonant production, $m^{\rm mod}_{\rm reslim}$, using the value of $\epsilon_{\rm res}$ at the end of resonant production. These two improved boundaries more accurately delineate the parameter space where resonant production suppresses non-resonant production and where significant resonance does not occur.

 The temperature $T_{\rm end}$, corresponding to the point where $F_{\rm res}$ has only one root, is obtained by setting the discriminant of Eq.~\eqref{eq:Fres} to zero. This yields
\begin{equation}
\label{eq:Tend}
    T_{\rm end} = \frac{\pi^2 m_s \sqrt{r_\alpha\cos2\theta}}{2\zeta(3)\mathcal{L}_\alpha} \simeq 36\textrm{ MeV}\left(\frac{m_s}{\textrm{keV}}\right) \left(\frac{\mathcal{L}_\alpha}{10^{-3}}\right)^{-1}.
\end{equation}
 The condition $T_{\rm end} > T_{\rm max}$ ensures that non-resonant production remains active after resonance ends. If instead $T_{\rm end} < T_{\rm max}$, the remaining lepton asymmetry continues to affect $F_{\rm res}$ and suppresses non-resonant sterile neutrino production.

For the new modified estimate of the upper mass limit, $m^{\rm mod}_{\rm reslim}$, above which resonant production does not significantly occur, the lepton asymmetry should be taken at its initial value, $\mathcal{L}_\alpha = \mathcal{L}_{\alpha,0}$, since without resonance the lepton number remains not depleted. In contrast, to estimate the lower mass limit, $m^{\rm mod}_{\rm non-res}$, above which non-resonant production can occur after the resonant phase ends, the depleted lepton number at $T_{\rm end}$, denoted $\mathcal{L}_\alpha(T_{\rm end})$, must be used. This is because resonant production partially depletes the lepton asymmetry before non-resonant production becomes active. The value of $\mathcal{L}_\alpha(T_{\rm end})$ depends on the mixing angle. In the parameter space shown in Fig.~\ref{fig:fres-fnon-1} and Fig.~\ref{fig:fres-fnon-2} ($1~{\rm keV}\leq m_s\leq 100{~\rm keV}$ and $10^{-13}\leq\sin^22\theta\lesssim 2\times10^{-9}$), our numerical results find that $\mathcal{L}_\alpha(T_{\rm end})$ ranges from approximately $0.97\mathcal{L}_{\alpha,0}$ to $0.15\mathcal{L}_{\alpha,0}$ depending on the mixing.

For our new estimate of the lower mass limit, $m^{\rm mod}_{\rm non-res}$, we consider $\mathcal{L}_\alpha(T_{\rm end}) \simeq 0.15 \mathcal{L}_{\alpha,0}$. This choice corresponds to a scenario with substantial resonant production, ensuring the formation of a sizable cool population. The condition $T_{\rm end} > T_{\rm max}$ then leads to the bound
\begin{equation}
\label{eq:mslower}
    m_s> m^{\rm mod}_{\rm non-res} \simeq 1.12\times10^3~ {\rm keV}\left[\frac{2\zeta(3)}{\pi^2(r_\alpha\cos2\theta)^{1/2}} \frac{\mathcal{L}(T_{\rm end})}{10^{-3}}\right]^{3/2}\simeq 0.29~{\rm keV}\left(\frac{\mathcal{L}_0}{10^{-3}}\right)^{3/2}. 
\end{equation} 

For sufficiently large masses, $m_s > m^{\rm mod}_{\rm reslim}$, the density and thermal potential terms in $F_{\rm res}$ become small, effectively preventing resonant production. To estimate this upper mass limit, we consider the largest possible value of $\epsilon$ for which a resonance can occur, corresponding to the maximum of the upper resonance branch, $\epsilon_{\rm res}^+$. Near the boundary where resonant production ceases, the lepton asymmetry remains nearly constant, so we approximate $\mathcal{L}_\alpha(T_{\rm end}) \simeq \mathcal{L}_{\alpha,0}$.
 By analytically solving $\partial \epsilon_{\rm res}^+ /\partial T = 0$, we find that the peak of the upper resonance branch occurs at temperature  $\sqrt{9/8}~T_{\rm end}$. At this point, the maximal value of $\epsilon_{\rm res}^+$ is 
\begin{equation}
\label{eq:erespeak}
    \epsilon_{\rm res}^+ (\sqrt{9/8}~T_{\rm end}) 
    \simeq 0.84 \frac{(\cos2\theta)^{1/2}m_s}{r_\alpha^{1/2} T_{\rm end}^3}~.
\end{equation}
Little resonant production occurs if $\epsilon_{\rm res}^+(\sqrt{9/8}~T_{\rm end}) \lesssim 0.1$, due to the suppression of the Fermi-Dirac distribution at low momentum, where the number density of active neutrinos is proportional to $\epsilon^2$. In this case, the resulting cool sterile neutrino population remains a subdominant contribution to DM, with an abundance bounded by $f_{\rm s,DM} \lesssim 4 \times 10^{-3} (m_s/\textrm{keV})( 10.75/g_*(t_{\rm prod}))$, where this limit assumes complete conversion of active neutrinos with $\epsilon < 0.1$.
 
Fig.~\ref{fig:msres} shows that  for a given lepton asymmetry $\mathcal{L}_0$, the shape of the two resonance branches $\epsilon_{\rm res}^{\pm}$ depends on the sterile neutrino mass $m_s$. In particular, the maximum of the upper resonance branch, $\epsilon_{\rm res}^+$, decreases with increasing mass approximately as $\propto m_s^{-2}$, consistent with Eqs.~\eqref{eq:erespeak} and~\eqref{eq:Tend}. By inverting Eq.~\eqref{eq:erespeak} for $m_s$ and requiring $\epsilon_{\rm res}^+(\sqrt{9/8}~T_{\rm end}) > 0.1$, we derive a necessary condition for significant resonant production
\begin{eqnarray}
\label{eq:msupper}
    m_s<m^{\rm mod}_{\rm reslim}\simeq 40.62~{\rm keV}\left(\frac{\mathcal{L}_0}{10^{-3}}\right)^{3/2}.
\end{eqnarray} 
This condition for significant resonant production is equivalent to requiring that the resonance curves intersect the momentum value $\epsilon = 0.1$ at least once.

Both Eq.~\eqref{eq:mslower} and Eq.~\eqref{eq:msupper} provide necessary conditions for the existence of two distinct sterile neutrino populations: one produced resonantly and the other non-resonantly. However, these conditions alone do not determine the relative abundances of each component. The fraction of DM composed of sterile neutrinos, $f_{\rm s,DM}$, depends on the parameters $m_s$, $\sin^2 2\theta$, and $\mathcal{L}_0$. To identify the regions of parameter space where significant two-population production occurs, we perform a numerical scan over these parameters and compute the density fraction of each population for each parameter set.

\begin{table}[t]
    \centering
    \begin{tabular}{|c|c|c|c|c|c|c|c|}
    \hline  
     Model & $m_s$(keV) & $\sin^22\theta(10^{-10})$ & $L_{e,0}(10^{-3})$ & $f_{\rm s,DM}^{\rm cool}$ & $f_{\rm s,DM}^{\rm warm}$&$\langle\epsilon_{\rm ref}\rangle_{\rm cool}$&$\langle\epsilon_{\rm ref}\rangle_{\rm warm}$\\
     \hline \hline
     1 & 4.30 & 1.10 & 2.98 & 0.86 & 0.10& 0.82&3.20\\
      \hline
      2 & 7.80 & 0.24 & 2.50 & 0.87 & 0.09& 1.01&3.25\\
      \hline
      3 & 7.14 & 0.32 & 2.60 & 0.88 & 0.10& 0.96&3.19\\
      \hline
      4 & 15.00 & 0.60 & 1.20 & 0.48 & 0.50& 0.86&2.17\\
       \hline
    \end{tabular}
    \caption{Model examples of sizable cool and warm populations of the same sterile neutrino species produced via flavor oscillations. The DM density fraction and the average temperature-scaled momentum at $T = 3$ MeV for each component, with the combined sterile neutrino abundance satisfying $f_{\rm s,DM} \simeq 1$ are shown.}
    \label{tab:fs=1eg}
    \end{table} 

Fig.~\ref{fig:fres-fnon-1} shows the regions of parameter space where both cool and warm sterile neutrino populations are produced with comparable abundances. The colored bands indicate where the ratio of number densities satisfies $0.1 < n_{\nu_s,\rm cool}/n_{\nu_s,\rm warm} < 10$, for an initial lepton asymmetry of $\mathcal{L}_0 = 10^{-3}$. The scan covers the range $1~\mathrm{keV} < m_s < 100~\mathrm{keV}$ and $10^{-13} < \sin^2 2\theta \lesssim 3 \times 10^{-9}$. In the green band, the cool population dominates, while in the yellow band, the warm population is dominant. Although in this example the correct DM abundance can be produced along the slanted black solid line (above which $f_{\rm s,DM} > 1$ and the gray region is therefore excluded), the resulting sterile neutrinos would constitute warm dark matter and are ruled out by Lyman-$\alpha$ forest and strong lensing constraints.

In Fig.~\ref{fig:fres-fnon-2} we show the results for a larger initial lepton asymmetry, $\mathcal{L}_0 = 4 \times 10^{-3}$. In this case, the cool population dominates over the warm component, with the $f_{\rm s,DM} = 1$ black line lying entirely within the green band for sterile neutrino masses in the range $7~\mathrm{keV} \lesssim m_s \lesssim 70~\mathrm{keV}$. This implies that sterile neutrinos in this parameter space would behave predominantly as CDM  and could account for all of DM, in the regions allowed by existing X-ray constraints.

In Tab.~\ref{tab:fs=1eg} we include a set of representative models that produce two sterile neutrino populations via neutrino oscillations. Here, we show for both the cool and warm sterile neutrino populations the corresponding fractions of DM, $f^{\rm cool}_{s, \rm DM}$ and $f^{\rm warm}_{s, \rm DM}$, as well as the average momentum at $T = 3$ MeV, $\langle\epsilon_{\rm ref}\rangle_{\rm cool}$ and $\langle\epsilon_{\rm ref}\rangle_{\rm warm}$. Here, $\langle\epsilon_{\rm ref}\rangle = \langle\epsilon(t)\rangle = \langle\epsilon_g\rangle \left[g_*(T=3~\mathrm{MeV})\right]^{1/3}$. For these parameters the combined density of the two components accounts for the total DM abundance, with sizable contributions from both the cool and warm populations. Using Eq.~\eqref{eq:massrelation} we account for  the warm population that is allowed by the Lyman-$\alpha$ bound.  The first two benchmark models correspond to two-population sterile neutrino DM scenarios that lie below current Nuclear Spectroscopic Telescope Array (NuSTAR) limits~\cite{Neronov:2016wdd,Perez:2016tcq,Ng:2019gch}, that may be probed by upcoming and ongoing X-ray missions such as XRISM~\cite{XRISMScienceTeam:2020rvx,Zhou:2024sto}. The third model represents a potential explanation of the 3.57 keV X-ray line~\cite{Bulbul:2014sua}. All three of these models involve significant resonant production, resulting in sufficiently cool sterile neutrino populations to evade current Lyman-$\alpha$ constraints while remaining consistent with X-ray bounds.
The fourth benchmark point corresponds to a heavier sterile neutrino that could account for the total DM abundance, with approximately equal contributions from the cool and warm components. This scenario satisfies Lyman-$\alpha$ constraints but is in tension with current X-ray limits.

\section{Gravitational Production: Primordial Black Hole Neutrinogenesis}
\label{sec:neutrinogenesis}

\subsection{Evaporating PBHs}

PBH sterile neutrinogenesis—the production of sterile neutrinos from Hawking evaporation of early Universe black holes—was studied in Refs.~\cite{Chen:2023lnj,Chen:2023tzd}, focusing on a single-mass monochromatic PBH population. Unlike oscillation-based sterile neutrino production, PBH neutrinogenesis is purely gravitational and independent of sterile-active mixing or other particle couplings, providing a nonthermal source of sterile neutrinos that is insensitive to traditional model parameters. Here, we first explore the combined impact of PBH evaporation and oscillation production on the sterile neutrino spectrum. We then extend the analysis to consider two distinct PBH populations with masses $M_1$ and $M_2$ ($M_1 < M_2$) and discuss how multiple evaporation channels affect sterile neutrino production and the resulting DM abundance.

To be concrete, we consider as a reference the extensively studied scenario in which PBHs are produced from the collapse of horizon-sized density perturbations. In this case, the PBH mass is related to the formation temperature by assuming that PBHs form with a mass equal to a fraction $\gamma \simeq 0.2$ of the horizon mass at formation, as is typical for near-critical collapse~(e.g.~\cite{Sasaki:2018dmp}) 
\begin{equation}
    T_{\rm form} \simeq 4.35\times10^{11}\textrm{ GeV} \left(\frac{10^8~{\rm g}}{M_{\rm PBH}}\right)^{1/2} \left(\frac{\gamma}{0.2}\right)^{1/2} \left(\frac{106.75}{g_*(T_{\rm form})}\right)^{1/4}~.
\end{equation}
However, our results can be readily generalized to PBHs formed through other mechanisms.

As the PBH evaporates, the rate of decay $dM/dt \propto 1/M^2$~\cite{Page:1976df} accelerates, resulting in a burst of radiation at the end of the PBH lifetime. Integrating this relation, the lifetime can be found to scale strongly with the initial mass,
\begin{equation}
\label{eq:PBHlifetime}
    \tau_{\rm evap} = 3.9\times10^{-4} ~{\rm s} \left(\frac{M_{\rm PBH}}{10^8~{\rm g}}\right)^3 \left(\frac{110}{g_H}\right)~.
\end{equation}
The number of degrees of freedom in Hawking radiation is taken as $g_H = 110$, accounting for all the 108 SM degrees of freedom plus the additional 2 from sterile neutrinos. If PBHs do not dominate the energy density before they evaporate, the Universe is assumed to remain radiation dominated until the standard matter-radiation equality epoch. In this case, the PBH lifetime given in Eq.~\eqref{eq:PBHlifetime} can be related to the temperature of the Standard Model plasma at the time of evaporation~\cite{Hooper:2019gtx} 
\begin{equation}
\label{eq:Tevap}
    T_{\rm evap} \simeq 43\textrm{ MeV} \left(\frac{10^8 ~{\rm g}}{M_{\rm PBH}}\right)^{3/2} \left(\frac{10.75}{g_*(T_{\rm evap})}\right)^{1/4}\left(\frac{g_H}{110}\right)^{1/2}~.
\end{equation}
Note that lighter PBHs evaporate earlier.

Alternatively, if PBHs temporarily dominate the energy density, their evaporation reheats the Universe to a temperature slightly higher than in the standard radiation-dominated scenario 
\begin{equation}
\label{eq:TRH}
    T_{\rm RH} \simeq 55\textrm{ MeV} \left(\frac{10^8 ~{\rm g}}{M_{\rm PBH}}\right)^{3/2} \left(\frac{10.75}{g_{*}(T_{\rm RH})}\right)^{1/4} \left(\frac{g_H}{110}\right)^{1/2}~.
\end{equation}
Since we typically consider parameters where $T_{\rm RH} > 100~\mathrm{GeV}$ in this section, the effective number of relativistic degrees of freedom remains approximately constant, with $g_{*}(T_{\rm RH}) = 106.75$. Therefore, we assume a fixed $g_{*}$ during production and present the results in terms of $\epsilon$ redshifted to $g_{*}(T_{\rm dc}) = 10.75$, rather than using $\epsilon_g$.

To avoid potential complications with BBN, we restrict our analysis to PBHs with masses $M \lesssim 4 \times 10^{8}$~g , corresponding to a reheating temperature $T_{\rm RH} > 5~\mathrm{MeV}$. The Hawking temperature of PBHs is given by
\begin{equation}
    T_{\rm PBH} = 1.06\times10^5 \textrm{ GeV} \left(\frac{10^8 ~{\rm g}}{M_{\rm PBH}}\right)~.
\end{equation}
The corresponding Hawking temperature is then high enough to emit the entire SM spectrum, as well as the full range of sterile neutrino masses considered in this work~$\textrm{keV} \lesssim m_s \lesssim \textrm{GeV}$. The average initial momentum of the particles produced via PBH sterile neutrinogenesis is significant and increases with PBH mass
\begin{equation} \label{eq:pbhmom}
    \left\langle\epsilon\right\rangle \simeq 6.3\frac{T_{\rm PBH}}{T_{\rm evap}} \left(\frac{10.75}{g_*(T_{\rm evap})}\right)^{1/3} = 1.5\times10^7 \left(\frac{M_{\rm PBH}}{10^8 ~{\rm g}}\right)^{1/2}\left(\frac{10.75}{g_*(T_{\rm evap})}\right)^{1/12}.
\end{equation}

From these relations, the lighter PBH population with mass $M_1$ forms and evaporates earlier, resulting in a higher reheating or evaporation temperature compared to the heavier population with mass $M_2$. For the remainder of this section, we label quantities with subscripts $1$ and $2$ to denote the sterile neutrino populations produced by the evaporation of the lighter and heavier PBHs, respectively.

\subsection{Neutrinogenesis and active-sterile oscillations}
 
Since sterile neutrinos are assumed to have non-zero mixing with active neutrinos, production via the DW mechanism occurs alongside PBH sterile neutrinogenesis. These two production channels yield sterile neutrinos with distinctly separated momentum spectra. The DW mechanism typically produces a modest average momentum of $\langle \epsilon \rangle \simeq 3.15~[g_{*}(T_{\rm max})/10.75]^{1/3}$, whereas PBH neutrinogenesis leads to a significantly hotter population with $\langle \epsilon \rangle$ given by Eq.~\eqref{eq:pbhmom}.
For sufficiently large mixing angles, the colder DW population can contribute a comparable density to the PBH-induced population~\cite{Chen:2023lnj,Chen:2023tzd}. However, keV–MeV scale sterile neutrinos produced via the DW mechanism are tightly constrained by Lyman-$\alpha$ forest data, strong lensing limits on warm DM, and X-ray bounds from radiative decays (see Sec.~\ref{sec:limits}). As a result, their contribution to the DM abundance is typically restricted to $\lesssim 10\%$ in this mass range~\cite{Gelmini:2019wfp}.
In contrast, PBH sterile neutrinogenesis is independent of the mixing angle and can evade X-ray constraints if the mixing is sufficiently small, while still producing the bulk of the DM~\cite{Chen:2023lnj,Chen:2023tzd}. To satisfy WDM constraints, the highly boosted sterile neutrinos from PBH evaporation must have masses $m_s \gtrsim \textrm{MeV}$.

For heavier sterile neutrinos with $m_s \gtrsim \textrm{MeV}$, non-resonant oscillations could in principle generate a second population. However, DW production in this mass range is strongly constrained by $\gamma$-ray observations from sterile neutrino decays~\cite{DeRomeri:2020wng}. We have verified that as a result of this any two-population scenario would necessarily involve only a vanishingly small fraction of sterile neutrinos produced via oscillations.
Nonetheless, in the presence of modified cosmologies, non-standard neutrino interactions, or other new physics, the sterile neutrino production rate via oscillations could be significantly enhanced~\cite{DeGouvea:2019wpf,Kelly:2020pcy,Chichiri:2021wvw}.

Additionally, PBH neutrinogenesis and DW production interact non-trivially when the PBH reheating temperature after an early matter-dominated phase, given by Eq.~\eqref{eq:TRH}, is lower than the temperature at which the DW production rate is maximized, Eq.~\eqref{eq:Tmax}. In this case, DW production is suppressed because it only begins after reheating, similar to what occurs in low-reheating temperature cosmologies (see e.g.~\cite{Gelmini:2004ah, Gelmini:2008fq, Rehagen:2014vna}).

The occupation number of high-momentum sterile neutrinos produced by PBH neutrinogenesis exceeds that of the equilibrium Fermi-Dirac distribution, which is Boltzmann suppressed and effectively zero at large $\epsilon$. This raises the possibility that, at large mixing angles, high-momentum sterile neutrinos could oscillate back into active neutrinos as described by Eq.~\eqref{eq:boltzmannp}, potentially depleting this sterile neutrino population. However, Ref.~\cite{Chen:2023tzd} showed that this effect is equivalent to thermalization and only becomes significant for mixing angles well within the region already excluded by existing X-ray constraints~\cite{Gelmini:2020duq}.

 \begin{table}[t]
 \centering
    \begin{tabular}{|c|c|c|c|c|c|c|c|}
    \hline
     Model & $m_s$~(keV) & $M_1~({\rm g})$ & $\beta_1$ & $f_{\rm s,1DM}$ & $M_2~({\rm g})$ & $\beta_2$&$f_{\rm s,2DM}$\\
    \hline \hline
        1&$1\times 10^{5}$ & $1$ &$2.5\times10^{-9}$&0.5&$66$&$3\times10^{-10}$&$0.5$\\
    \hline
        2&$4\times10^5$ &$40$& $1.2\times10^{-9}$& 0.5 & $3.2\times10^3$ & $1.4\times10^{-10}$  & 0.5 \\
    \hline
       3&$2\times10^6$ &$3\times10^{3}$& $2.8\times10^{-11}$& 0.5 & $1.6\times10^{5}$ & $3.8\times10^{-12}$  & 0.5 \\
       \hline
    \end{tabular}
    \caption{Benchmark models of PBH sterile neutrinogenesis from two PBH populations resulting in a mixed population DM scenario, where cold and warm components each contribute approximately half of the total DM density. The lighter PBHs with mass $M_1$ primarily produce the cold component, while the heavier PBHs with mass $M_2$ produce the warm component.}
    \label{tab:doubleneutrino}
\end{table}

\subsection{Spectrum with sub-dominant PBH populations}
\label{ssec:nodom}

We analyze the interplay between two monochromatic populations of light PBHs, each of which emits a corresponding population of sterile neutrinos through evaporation. We present 
in Tab.~\ref{tab:doubleneutrino} several characteristic models of such scenario, including their resulting contributions to sterile neutrino DM abundance.
If a PBH matter-dominated phase occurs, the scale factor evolves differently compared to standard radiation domination. As a result, both the momentum distribution and the abundance of the emitted sterile neutrinos are modified relative to the case where the PBHs evaporate without dominating the energy density of the Universe. We first consider this non-dominating case, which requires that the initial PBH energy fractions at formation, $\beta_1$ and $\beta_2$, are sufficiently small.

Since PBHs scale as matter, their energy density grows relative to radiation as the Universe expands. To avoid PBH matter-domination, the PBH energy density at the time of evaporation must remain subdominant compared to the radiation density 
\begin{equation}
    \beta_1 \frac{a_{\rm evap,1}}{a_{\rm form,1}} + \beta_2 \frac{a_{\rm evap,1}}{a_{\rm form,2}} <1,~~~{\rm and}~~~ \beta_2 \frac{a_{\rm evap,2}}{a_{\rm form,2}} < 1~.
\end{equation} 
Here, $a_{\rm form,i}$ and $a_{\rm evap,i}$ denote the scale factors at formation and evaporation of each PBH population, with $i=1,2$.
In the absence of a PBH matter-dominated phase, the initial PBH fractions $\beta_i$ are related to the PBH fractions at evaporation $f_{\rm evap,i}$  by
\begin{equation}
    f_{\rm evap,i} \simeq \beta_i \left(\frac{g_*(T_{\rm form,i})}{g_*(T_{\rm evap,i})}\right)^{1/3}\left(\frac{T_{\rm form,i}}{T_{\rm evap,i}}\right)~.
\end{equation}
The number density of sterile neutrinos at the time of PBH evaporation in the radiation-dominated case is given by~\cite{Chen:2023lnj,Chen:2023tzd}
\begin{equation}
\label{eq:numtevap}
n_{s,i}(T_{\rm evap,i}) = \frac{f_{\rm evap,i}}{10395}   \frac{\pi^2 g_*(T_{\rm evap,i})  T_{\rm evap,i}^4}{ T_{\rm PBH,i} }~.
\end{equation}
After evaporation, the scale factor evolves under radiation domination so that the present-day number density is given by
\begin{equation}
\label{eq:ns0}
n_{s,i0} = n_{s,i}(T_{\rm evap,i}) \left(\frac{a_0}{a_{\rm evap,i}}\right)^3
\simeq \frac{f_{\rm evap,i}}{10395}  \frac{\pi^2 g_{s*}(T_0)  T_{\rm evap,i}  T_0^3}{ T_{\rm PBH,i} }~,
\end{equation}
where the subscript $0$ denotes present-day quantities, with $T_0 = 2.73$ K as the current photon temperature and $g_{s*}(T_0) = 3.9$ as the present-day effective number of entropy degrees of freedom.
As discussed in Sec.~\ref{ssec:dom}, if PBHs dominate the energy density before evaporation, both the relation between $\beta_i$ and $f_{\rm evap,i}$ and the post-evaporation scale factor evolution are modified.

\begin{figure}[t]
    \centering
    \includegraphics[width=0.49\linewidth]{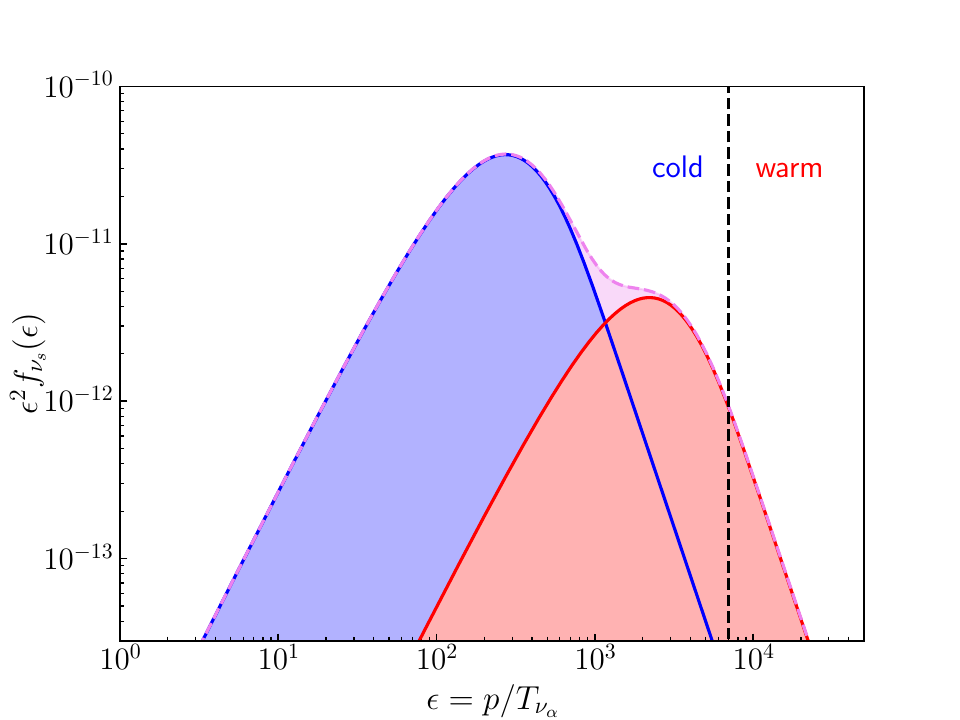}
    \caption{ 
    Momentum distribution as function of $\epsilon=p/T_{\nu_s}$ evaluated at the reference temperature $T_{\rm ref} = 3~\mathrm{MeV}$ for $m_s = 1 \times 10^{5}~\mathrm{keV}$ sterile neutrinos from PBH sterile neutrinogenesis with two monochromatic PBH populations of masses $M_1 = 1~\mathrm{g}$ (blue) and $M_2 = 66~\mathrm{g}$ (red) corresponding to Model 1 in Tab.~\ref{tab:doubleneutrino}. The combined distribution is also shown (dashed light violet). 
    This model has 
    approximately $50\%$ cold DM and $50\%$ warm DM. 
    The vertical black dashed line indicates the average scaled momentum boundary between cold and warm DM populations, corresponding to a thermal relic mass of $m_{\rm therm} = 5.7~\mathrm{keV}$,  
    with $\epsilon$ evaluated at 
    $T_{\rm ref} = 3~\mathrm{MeV}$.
    } 
    \label{fig:twoneutrinogenesis}
\end{figure}

Despite the highly energetic spectrum produced by PBH neutrinogenesis, moderately heavy sterile neutrinos with masses $m_s > \textrm{keV}$ become non-relativistic by the present day, and those with $m_s \gtrsim \textrm{MeV}$ can behave as cold DM (see Ref.~\cite{Chen:2023tzd}). Using the non-relativistic relation $\rho_s = n_{s,0} m_s$ together with Eq.~\eqref{eq:ns0}, the DM fraction from PBH evaporation $f_{\rm s,DM} = \rho_s/\rho_{\rm DM}$ is  
\begin{equation}
    f_{\rm s,iDM} \simeq ~2\times10^{-6} \sum_i   f_{\rm evap,i} \left(\frac{m_s}{\textrm{keV}}\right)\left(\frac{10^8~{\rm g}}{M_{\rm PBH,i}}\right)^{1/2}\left(\frac{10.75}{g_*(T_{\rm evap,i})}\right)^{1/4}.
\end{equation} 
In the case where PBHs do not dominate the energy density, the relic momentum distributions from the two PBH populations simply add, resulting in two distinct momentum peaks at $\langle\epsilon\rangle_i$. This two-population PBH neutrinogenesis scenario can produce either a mixed DM distribution, with one cold and one warm component, or a spectrum consisting of two distinct warm components with different characteristic momenta.
 
In Fig.~\ref{fig:twoneutrinogenesis}, considering model 1 of Tab.~\ref{tab:doubleneutrino}, we illustrate sterile neutrino momentum distributions from two monochromatic PBH populations evaporating into sterile neutrinos with distinct average momenta. The resulting spectrum includes both cold and  warm DM components, yielding approximately equal amounts of cold and warm DM  accounting for the entire DM abundance, which could have implications for structure formation. The average momentum of each component can be compared to the vertical dashed black line in the figure, which separates warm and hot DM populations. Given $m_s$, each average momentum corresponds to a thermal relic DM mass $m_{\rm therm}$, determined using Eq.~\eqref{eq:massrelation} to convert thermal relic constraints from Refs.~\cite{Baur:2017stq,Gilman:2019nap,Zelko:2022tgf,Garcia-Gallego:2025kiw} into equivalent limits for sterile neutrinos~\cite{Viel:2005qj,Gelmini:2019wfp}.  
 We adopt $m_{\rm therm} \simeq 5.7~\textrm{keV}$ as the boundary between cold and warm DM~\cite{Garcia-Gallego:2025kiw}, which yields an average scaled momentum, indicated for $m_s = 1\times 10^{5}$ keV by the dashed black vertical  line in Fig.~\ref{fig:twoneutrinogenesis}. 
  Thermal relics with mass above this limit (i.e. with smaller average momentum for a given value of $m_s$) are
  unconstrained by current Lyman-$\alpha$ data, while strong lensing further disfavors thermal relics with 
  mass below $\simeq 5$ keV (corresponding to larger average momentum in the figure).
  
  As explained in Sec.~\ref{sec:limits} we use Eq.~\eqref{eq:massrelation} to limit the contribution of a WDM population, since for a given $m_s$, the computed average scaled momentum of a population corresponds to a specific $m_{\rm therm}$ that can then be matched with the allowed DM fraction.
  
The sterile neutrino momentum distributions shown in Fig.~\ref{fig:twoneutrinogenesis} are computed numerically using the approximation of Ref.~\cite{Baldes:2020nuv}
\begin{equation}
\epsilon^2 f_{\nu_{s,i}}(\epsilon) \simeq \left(\frac{T_{\rm PBH,i}}{p}\right)^{3} \int_{0}^{p/T_{\rm PBH,i}} \frac{y^4}{e^y + 1} dy~,
\end{equation}
with normalization fixed by the number density in Eq.~\eqref{eq:numtevap}.

\subsection{Spectrum with PBHs that matter-dominate before evaporating}
\label{ssec:dom}

Matter domination by either PBH population alters both the present-day number density and the momentum spectrum of the sterile neutrinos produced via evaporation. Here, we consider the effect of matter domination from each PBH population separately. While we do not explicitly analyze the case where both populations induce consecutive or overlapping matter-dominated eras, such scenarios can be approximated by appropriately combining the individual effects discussed here.

First, we consider the scenario where the lighter PBH population with mass $M_1$ evaporates while remaining subdominant, followed by a period where the heavier PBH population with mass $M_2$ comes to dominate the energy density of the Universe. The conditions for this scenario are 
\begin{equation}
\label{eq:M2domcond}
\beta_1 \frac{a_{\rm evap,1}}{a_{\rm form,1}} + \beta_2 \frac{a_{\rm evap,1}}{a_{\rm form,2}} < 1, \qquad \text{and} \qquad \beta_2 \frac{a_{\rm RH,2}}{a_{\rm form,2}} > 1~.
\end{equation}
Here, $\beta_1$ and $\beta_2$ are the initial PBH fractions of the total energy density at formation, and $a_{\rm evap,i}$ and $a_{\rm form,i}$ denote the scale factors at evaporation and formation for each PBH population, respectively. The scale factor $a_{\rm RH,2}$ corresponds to the reheating time after the evaporation of the $M_2$ PBHs.
In this case, the sterile neutrinos produced by the $M_1$ population ($\nu_{s,1}$) experience additional redshifting due to the subsequent matter-dominated era induced by the $M_2$ PBHs. As a result, the ratio $a_0/a_{\rm evap,1}$ is effectively increased, leading to a suppression of both the present-day number density and the average momentum of the $\nu_{s,1}$ population. In contrast, the sterile neutrinos produced by the $M_2$ PBHs ($\nu_{s,2}$) are emitted at reheating and redshift as usual under radiation domination.

The redshift evolution of the sterile neutrino momentum can be broken into three stages. From $M _1$ evaporation to $M_2$ matter-domination  the Universe is radiation dominated  and the scale factor evolves as
\begin{equation}
\frac{a_{\rm tr,2}}{a_{\rm evap,1}} = \left(\frac{g_*(T_{\rm evap,1})}{g_*(T_{\rm tr,2})}\right)^{1/3} \frac{T_{\rm evap,1}}{T_{\rm tr,2}}~,
\end{equation}
where  $T_{\rm tr,2}$ denotes the temperature at which the Universe transitions from radiation domination to $M_2$ PBH matter-domination.
During the matter-dominated era between $T_{\rm tr,2}$ and the evaporation of the $M_2$ PBHs  the scale factor grows as
\begin{equation}
\frac{a_{\rm RH,2}}{a_{\rm tr,2}} = \frac{T_{\rm tr,2}}{T_{\rm RH,2}}~,
\end{equation}
where $T_{\rm RH,2}$ is the reheating temperature corresponding to $M_2$ evaporation.
After reheating, the Universe returns to radiation domination and the scale factor evolves to the present day as
\begin{equation}
\frac{a_0}{a_{\rm RH,2}} = \left(\frac{g_*(T_0)}{g_*(T_{\rm RH,2})}\right)^{1/3} \frac{T_0}{T_{\rm RH,2}}~.
\end{equation} 
  
The scale factor growth during the intermediate period, corresponding to the $M_2$ PBH matter-dominated era, can be related to the lifetime of the $M_2$ PBHs. Since this phase is matter dominated, the Hubble expansion behaves as $H(a) \propto a^{-3/2}$. The lifetime of the $M_2$ PBHs is given by 
\begin{align}
    \tau_2 \simeq  \frac{1}{2 H(T_{\rm tr,2})} - t_{\rm form,2}+\frac{1}{H(T_{\rm tr,2})}\int_{a_{\rm tr,2}}^{a_{\rm evap,2}} \frac{da}{a} \left(\frac{a}{a_{\rm tr,2}}\right)^{3/2} \simeq \frac{2}{3H_{T_{\rm tr,2}}} \frac{a_{\rm RH,2}}{a_{\rm tr,2}}~,
\end{align}
where  $H(T_{\rm tr,2})= 1/(2 t_{\rm tr,2})$ is the Hubble rate at the end of the preceding radiation dominated period, during which $H(T) = \sqrt{8\pi^3 g_*(T)/90}~ T^2/M_{\rm pl}$. The total redshift factor affecting the $\nu_s$ emitted by the $M_1$ PBHs is then
\begin{equation}
\label{eq:m2domeffect}
    \frac{a_0}{a_{\rm evap,1}} = \left(\frac{g_*(T_{\rm evap,1})}{g_*(T_0)}\right)^{1/3} \frac{T_{\rm evap,1}}{T_0} \left(\frac{g_*(T_{\rm evap,2})}{g_*(T_{\rm tr,2})}\right)^{1/3} \frac{T_{\rm evap,2}}{T_{\rm tr,2}} \left(\frac{3}{2}\tau_2 H(T_{\rm tr,2})\right)^{2/3}~.
\end{equation}

The period of $M_2$ PBH matter domination increases the overall redshift compared to the radiation-dominated case with
\begin{equation}
\left(\frac{a_0}{a_{\rm evap,1}}\right)_{\rm RD} = \left(\frac{g(T_{\rm evap,1})}{g_(T_0)}\right)^{1/3} \frac{T_{\rm evap,1}}{T_0},
\end{equation}
with the subscript $\rm RD$ denoting the radiation-dominated scenario without PBH domination.
The average momentum of sterile neutrinos produced by the $M_1$ PBHs is reduced by a factor of $(a_0/a_{\rm evap,1})/ \left(a_0/a_{\rm evap,1}\right)_{\rm RD} $
and the number density is suppressed by the cube of this factor.
According to Eq.~\eqref{eq:M2domcond}, a larger initial fraction $\beta_2$ of $M_2$ PBHs leads to a longer matter-dominated era and consequently stronger redshift effects.

A different effect arises if the $M_1$ PBHs dominate the Universe instead of the $M_2$ PBHs. Since the $M_1$ population evaporates first, there is no suppression of the $M_2$ sterile neutrino momenta due to subsequent redshifting. However, the number density of $M_2$ PBHs before evaporation is effectively reduced, resulting in decreased sterile neutrino production from the $M_2$ population.
In the case where the Universe remains radiation dominated at all times (see Sec.~\ref{ssec:nodom}), the energy density fraction of $M_2$ PBHs grows relative to radiation, leading to a PBH fraction at evaporation given by
\begin{equation}
\label{eq:fevap2nodom}
    f_{\rm evap,2} = \beta_2 \frac{a_{\rm evap,2}}{a_{\rm form,2}}  
    \simeq \beta_2 \left(\frac{g_*(T_{\rm form,2})}{g_*(T_{\rm evap,2})}\right)^{1/3} \frac{T_{\rm form,2}}{T_{\rm evap,2}}~.
\end{equation}

In contrast, when the $M_1$ PBHs dominate the Universe, the density fraction of the $M_2$ PBHs becomes effectively frozen during the $M_1$-dominated era at the value
\begin{equation}
    \frac{\rho_2}{\rho_{\rm tot}} = \frac{\beta_2}{\beta_2 + \beta_1 \left(\dfrac{a_{\rm form,2}}{a_{\rm form,1}}\right)}~,
\end{equation}
until the $M_1$ PBHs evaporate and reheat the Universe back to radiation domination. After reheating, the $M_2$ PBH density fraction resumes its growth relative to radiation, and at the time of $M_2$ evaporation the final fraction is
\begin{equation}
\label{eq:fevap2dom}
    f_{\rm evap,2} = \frac{\beta_2}{\beta_2 + \beta_1 \left(\dfrac{g_*(T_{\rm form,1}}{g_*(T_{\rm form,2}}\right)^{1/3}\left(\dfrac{T_{\rm form,1}}{T_{\rm form,2}}\right)} \left(\frac{g_*(T_{\rm RH,1})}{g_*(T_{\rm evap,2})}\right)^{1/3}\frac{T_{\rm RH,1}}{T_{\rm evap,2}}~.
\end{equation}
Comparing Eq.~\eqref{eq:fevap2nodom} and Eq.~\eqref{eq:fevap2dom}, 
we observe that $M_1$ PBH domination suppresses $f_{\rm evap,2}$, the fraction of $M_2$ PBHs at evaporation. Consequently, the abundance of sterile neutrinos emitted by the $M_2$ population is also reduced.
With appropriate modifications, the redshift effect from $M_2$ domination (see Eq.~\eqref{eq:m2domeffect}) and the dilution of $M_2$ PBHs from $M_1$ domination (see Eq.~\eqref{eq:fevap2dom}) can be combined to evaluate scenarios where both PBH populations impact the sterile neutrino abundance.

In addition to affecting small-scale structure formation through its two-peaked momentum spectrum, this scenario with several PBH populations could also produce a coincident multi-peaked gravitational wave spectrum. The evaporation of PBHs can trigger a rapid transition from matter domination to radiation domination, generating gravitational waves in the process~(e.g.~\cite{Inomata:2020lmk,Papanikolaou:2020qtd,Domenech:2020ssp,Domenech:2021wkk}). Here, the gravitational wave signals from the lighter $M_1$ PBHs, which evaporate first, may be partially suppressed by the subsequent matter domination of the heavier $M_2$ PBHs.
 
\section{Production from Particle Decays}
\label{sec:heavydecay}
 
Sterile neutrinos can also be produced via the decay of heavy particles, such as singlet Higgs bosons or inflatons. This mechanism differs from active-sterile oscillations, which depend on mixing angles and thermal plasma interactions, and from PBH neutrinogenesis that is purely gravitational and model-independent. In contrast, decay production relies on direct couplings to heavy fields and typically yields a colder, non-thermal sterile neutrino spectrum with average momenta set by the parent particle’s mass and decay dynamics. Previous studies, such as Ref.~\cite{Abazajian:2019ejt}, considered scenarios combining Higgs or GUT-scale particle decays with DW oscillations. These mixed production channels can produce sterile neutrinos with characteristic momenta spanning $\epsilon \sim 0.2$–$3$, resulting in an admixture of cold and warm DM components. Such spectra can help resolve small-scale structure tensions by suppressing the formation of dwarf galaxies while avoiding Lyman-$\alpha$ and strong lensing constraints.

Here, we go beyond previous treatments by explicitly tracking the full momentum distributions of sterile neutrinos produced via heavy particle decays combined with oscillations, using numerical Boltzmann evolution. We systematically map the parameter space where both decay and oscillation contributions coexist, quantify their respective fractions, and assess the cosmological and X-ray constraints. This allows us to identify viable mixed-phase DM scenarios where sterile neutrinos from decay dominate the cold component, while a subdominant warmer population arises from DW production.

A notable scenario where this can be realized is $\nu$MSM extension with an inflaton, proposed in Ref.~\cite{Shaposhnikov:2006xi}. The corresponding Lagrangian is given by
\begin{equation}
    \mathcal{L} \supset \mathcal{L}_{\nu \rm MSM} + \frac{1}{2}\left(\partial_\mu \chi\right)^2 - \frac{y_I}{2} \bar{N}_I^c N_I \chi + h.c. - V(\Phi,\chi)~,
\end{equation}
where $\mathcal{L}_{\nu\text{MSM}}$ contains the usual sterile neutrino-Higgs coupling, with the Higgs and Majorana mass terms are neglected for simplicity. The additional terms describe the kinetic and potential contributions of a real scalar inflaton $\chi$, together with its Yukawa coupling to sterile neutrinos through a constant coupling $y_I$.

Due to possible mixing interactions between the inflaton field $\chi$ of mass $m_{\chi}$ and the SM Higgs $\Phi$  the inflaton can thermalize with the plasma and remain in equilibrium until temperatures $T \lesssim m_\chi$.  This process produces a colder sterile neutrino momentum spectrum compared to the standard DW mechanism. The average momentum is approximately 
\begin{equation}
\langle\epsilon\rangle = 2.45\left(\dfrac{10.75}{Sg_{*}\left(m_\chi/3\right)}\right)^{1/3}
\end{equation}
where $g_{*}(m_\chi/3)$ is the effective number of relativistic degrees of freedom at the decay temperature, and $S \sim 1$–$2$ accounts for possible entropy dilution from the late decay of heavier sterile neutrino species. This colder spectrum complements the Fermi-Dirac-like distribution from DW production and helps alleviate small-scale structure constraints.
The sterile neutrino DM density produced from inflaton decay is given by~\cite{Shaposhnikov:2006xi}
\begin{equation}
    \Omega_s = \frac{\rho_s}{\rho_c} \simeq 8.6~ \frac{m_s}{m_\chi}\left(\frac{y_I}{10^{-12}}\right)^2\left(\frac{10.75}{g_*(m_\chi/3)}\right)^{2} \left(\frac{2}{S}\right)~,
\end{equation}
where $\rho_c$ is the present critical density of the Universe. 

The Higgs singlet decay model of Ref.~\cite{Kusenko:2006rh, Petraki:2007gq,Abazajian:2019ejt} is analogous to the inflaton decay scenario described above, but with $\chi$ interpreted as an $SU(2)$ singlet Higgs boson rather than the inflaton. In this case, the typical mass of $\chi$ is $\mathcal{O}(100)$ GeV, compared to $\mathcal{O}(1)$ GeV in the inflaton scenario. As a result, the decays occur at higher temperatures with a larger effective number of relativistic degrees of freedom, $g_{*}(m_\chi/3)$. Combined with an additional entropy dilution factor of $S \sim 2$, this leads to a colder sterile neutrino spectrum with an average momentum of $\langle\epsilon\rangle \simeq 0.9$.

  \begin{table}[t]
    \bigskip
\centering 
\begin{tabular}{|c|c|c|c|c|c|c|c|}  
\hline 
Model & $m_s$(keV) & $m_\chi$(GeV) & $y_I$ & $\langle\epsilon\rangle_{\rm decay}$ & $f_{\rm s, DM}^{\rm dec}$ & $\sin^2 2\theta$  & $f_{\rm s,DM}^{\rm osc}$  \\
\hline\hline
1 & $7.1$ & 250 & $2\times10^{-8}$ & 0.9 & $\simeq 1$ & $5\times10^{-11}$ & $4.2\times10^{-3}$ \\
\hline
2 & $7.1$ & 280 & $2\times10^{-8}$ & 0.9 & $0.9$ & $1.8\times10^{-9}$ & $0.1$ \\
\hline
3 & $8.1$ & 600 & $3\times10^{-8}$ & 0.9 & $\simeq 1$ & $6.1\times10^{-12}$ & $6.5\times10^{-4}$ \\
\hline
4 & $8.1$ & 670 & $3\times10^{-8}$ & 0.9 & $0.9$ & $1.2\times10^{-9}$ & $0.1$ \\
\hline
\end{tabular}
\caption{Singlet heavy scalar of mass $m_{\chi}$ and sterile neutrino parameters leading to a mixed DM population considering their interaction $y_I$. The dominant component $f_{\rm s, DM}^{\rm dec}$ with average momentum $\langle\epsilon\rangle_{\rm decay}$ is produced from scalar decays and a subdominant contribution $f_{\rm s,DM}^{\rm osc}$ constituting $\lesssim 10\%$ from DW production, including both $\nu_s$ and $\bar{\nu}_s$. These models are in slight tension with Lyman-$\alpha$ forest constraints, and model 2 and 4 are additionally constrained by current X-ray limits.
}
\label{tab:decaydw}
\end{table}

\subsection{Decays and active-sterile oscillations} 

Since a mixing between sterile and active neutrinos is generally assumed, an additional population of sterile neutrinos is inevitably produced via the DW mechanism. Typically, the term $f_{\nu_s}$ on the right-hand side of Eq.~\eqref{eq:boltzmannp} is negligible, allowing the abundances of different production channels to simply add together. This approximation was employed in Ref.~\cite{Merle:2014xpa}, where decay production was combined with DW production, and has been shown to be accurate to within a few percent~\cite{Merle:2015vzu}.
For intermediate values of the coupling constant $y_I$, it is possible to obtain two sterile neutrino populations from decay alone. In this scenario, one population originates from the decay of thermalized (equilibrated) scalars, while the other comes from the decay of frozen-out, out-of-equilibrium scalars resulting in comparable abundances but with a large difference in momentum distributions~\cite{Merle:2015oja}.

\begin{figure}[t]
    \centering
    \includegraphics[width=0.49\linewidth]{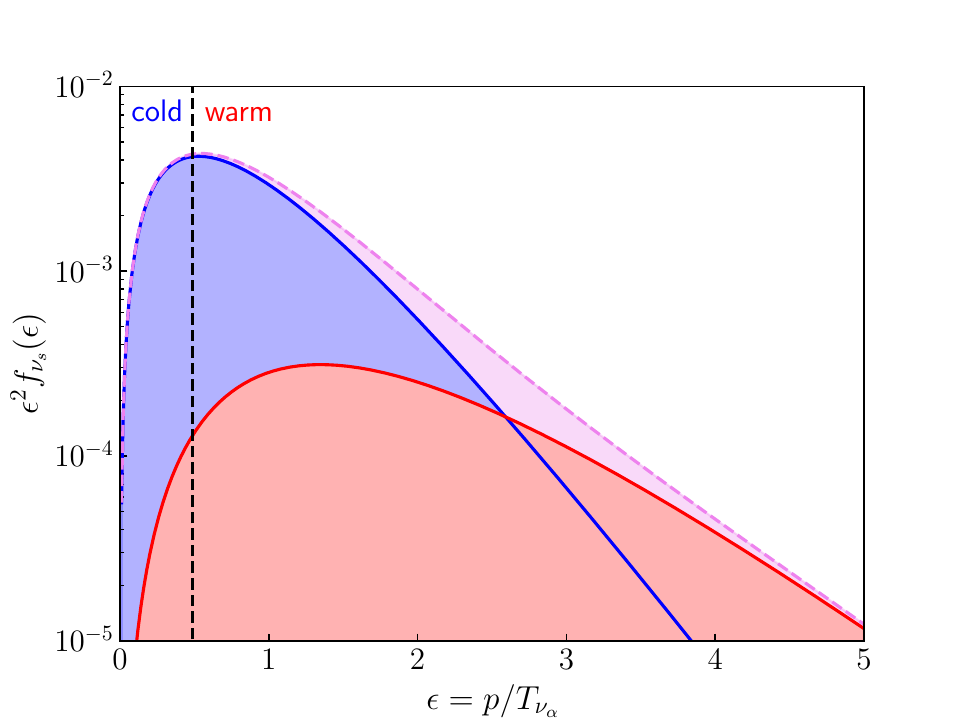}
    \caption{Two-component sterile neutrino population consisting of a colder component from heavy singlet decays (blue) and a warmer component from oscillations via the DW mechanism (red) for $m_s = 7.1$ keV, corresponding to model 2 of Tab.~\ref{tab:decaydw}. The dashed light violet line shows the combined momentum distribution. The vertical dashed black line corresponds to the average scaled momentum for $m_{\rm therm} = 5.7$ keV that we take to separate cold and warm DMpopulations.  
    }
    \label{fig:decaydw}
\end{figure}

For sufficiently small couplings $y_I$, only a single sterile neutrino population from out-of-equilibrium decays is produced, and this is the case we consider here in combination with DW production.
In Tab.~\ref{tab:decaydw}, we present four benchmark models. In model 1 and 3, sterile neutrinos from heavy singlet decays account for approximately $99\%$ of the DM in the form of a cool component (in slight tension with Lyman-$\alpha$ bounds), while a small warm DM component arises from DW production. The model with $m_s = 7.1$ keV could potentially account for the controversial $3.5$ keV X-ray line signal claimed in Ref.~\cite{Bulbul:2014sua,Boyarsky:2014jta}, whereas the $m_s = 8.1$ keV scenario lies at the edge of current X-ray constraints and could be tested by future X-ray observations such as from XRISM~\cite{Zhou:2024sto} (see Sec.~\ref{sec:limits}).
For higher mixing angles, the DW mechanism can produce up to $\sim10\%$ of DM in the form of warm components without violating Lyman-$\alpha$ limits (e.g. scenario 2 and 4). However, for the sterile neutrino masses considered in Tab.~\ref{tab:decaydw}, this region of parameter space is already constrained by current X-ray observations. 
In Fig.~\ref{fig:decaydw}, we show the mixed heavy singlet decay and DW-produced sterile neutrino momentum distribution corresponding to model 2. This case is in slight tension with the 2$\sigma$ Lyman-$\alpha$ bound~\cite{Garcia-Gallego:2025kiw}, as indicated by the position of the dashed line separating cold and warm DM components.

Other combinations of these production mechanisms could exhibit interesting interplay. For example, in the first GUT scenario discussed in Ref.~\cite{Abazajian:2019ejt}, DW production can be suppressed if decays produce a thermal population of sterile neutrinos that effectively block additional production through active-sterile oscillations. 

\begin{table}[t]
\centering 
\begin{tabular}{|c|c|c|c|c|c|c|c|c|}  
\hline 
Model & $m_s$(keV) & $m_\chi$(GeV) & $y_I$ & $\langle\epsilon\rangle_{\rm decay}$ & $f_{\rm s, DM}^{\rm dec}$ & $M_{\rm PBH}$(g) & $f_{\rm evap}$ & $f_{\rm s, DM}^{\rm PBH}$  \\
\hline \hline
1 & $20$ & 10 & $9.2\times10^{-10}$ & 0.99 & $0.9$ & 2 & 1 & 0.1 \\
\hline
2 & $40$ & 0.3 & $3.7\times10^{-11}$ & 1.64 & $0.9$ & 10 & 1 & 0.1 \\
\hline
3 & $4\times10^3$ & 600 & $4.7\times10^{-10}$ & 0.9 & $0.5$ & 0.11 & $0.0037$ & $0.5$ \\
\hline
\end{tabular}
\caption{Models of singlet heavy boson decays and PBH neutrinogenesis production yielding a mixed sterile neutrino DM population. In model 1 and 2, the spectrum consists of predominantly cold DM from singlet boson decay, with approximately $10\%$ warm DM from PBH neutrinogenesis. In model 3, we show approximately $50\%$–$50\%$ mix of cold DM from singlet decays and warm DM from PBH neutrinogenesis, corresponding to a sterile neutrino with mass of order $\mathcal{O}(\mathrm{MeV})$.
}
\label{tab:pbhdecay}
\end{table}

\subsection{PBH neutrinogenesis and decay production}

When decay production is paired with PBH neutrinogenesis, a prolonged early matter-dominated phase can effectively suppress any pre-existing sterile neutrino population, mimicking a low-reheating pre-BBN cosmology. For example, a reheating temperature of $T_{\rm RH} < 100~\mathrm{MeV}$, typical for PBHs with mass $M \gtrsim 10^8~\mathrm{g}$, would prevent a thermal population of heavy particles from entering equilibrium. In this case, only a small number of heavy particles frozen-in after PBH reheating could subsequently decay into sterile neutrinos, reducing their abundance.

\begin{figure}
    \centering
    \includegraphics[width=0.49\linewidth]{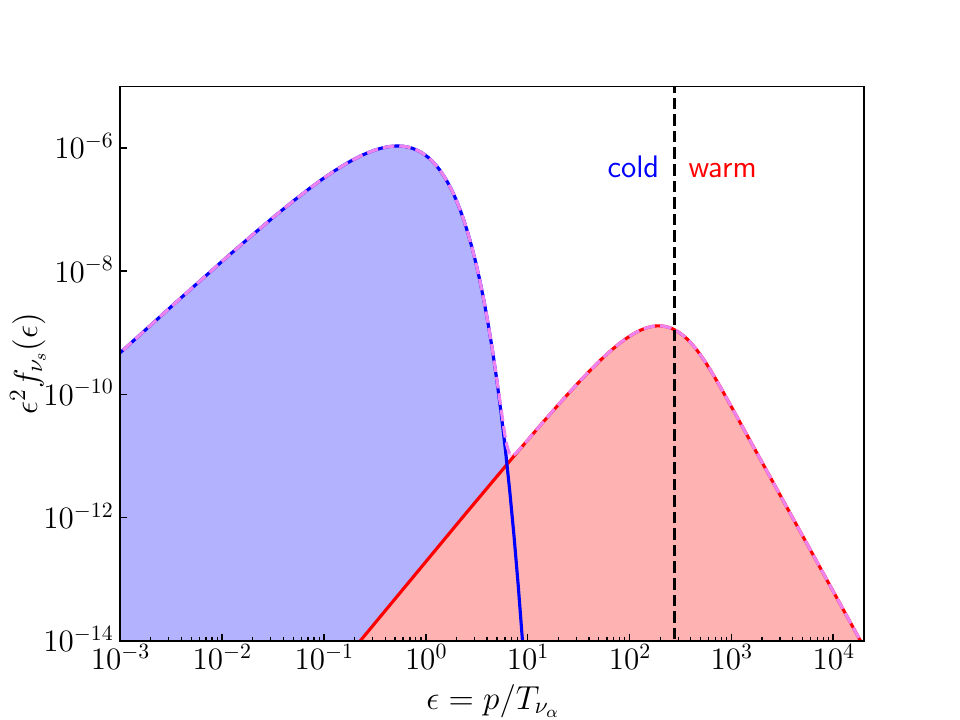}
    \caption{Momentum density as function of 
    $\epsilon = p/T_{\nu_s}$ at reference $T=3$ MeV for
    two-component sterile neutrino population consisting of approximately $50\%$ colder $\nu_s$ from heavy singlet boson decays (blue) and $50\%$ warmer $\nu_s$ from PBH neutrinogenesis (red), for $m_s = 4$ MeV, corresponding to model 3 of Tab.~\ref{tab:pbhdecay}. The combined population indicated with light violet dashed line closely follows the red and blue solid line. 
    The vertical dashed black line is the average scaled momentum that corresponds to $m_{\rm therm} = 5.7$ keV, separating cold and warm DM components.
    }
    \label{fig:pbhdecay}
\end{figure}

If the reheating temperature is sufficiently high or PBHs remain subdominant, the sterile neutrinos from decay and PBH production simply add together. In Tab.~\ref{tab:pbhdecay}, we present models that yield a majority cold DM component from singlet boson decay and approximately $10\%$ of warm DM from PBH neutrinogenesis. An example of this two-population scenario is shown in Fig.~\ref{fig:pbhdecay}, where the two momentum peaks are widely separated. Although the distributions appear to differ by orders of magnitude in $\epsilon^2 f_{\nu_s}$, the number density scales as $n_{\nu_s} \propto \epsilon^3 f_{\nu_s}$, introducing an additional factor of $\epsilon$. Consequently, the DM density fractions $f_{\rm s,DM}$ between the two populations are in an approximate ratio of $0.9 : 0.1$. 

\section{Cosmological, Astrophysical Constraints and Observables}
\label{sec:limits}

\begin{figure}[t]
    \centering
    \includegraphics[width=0.49\linewidth]{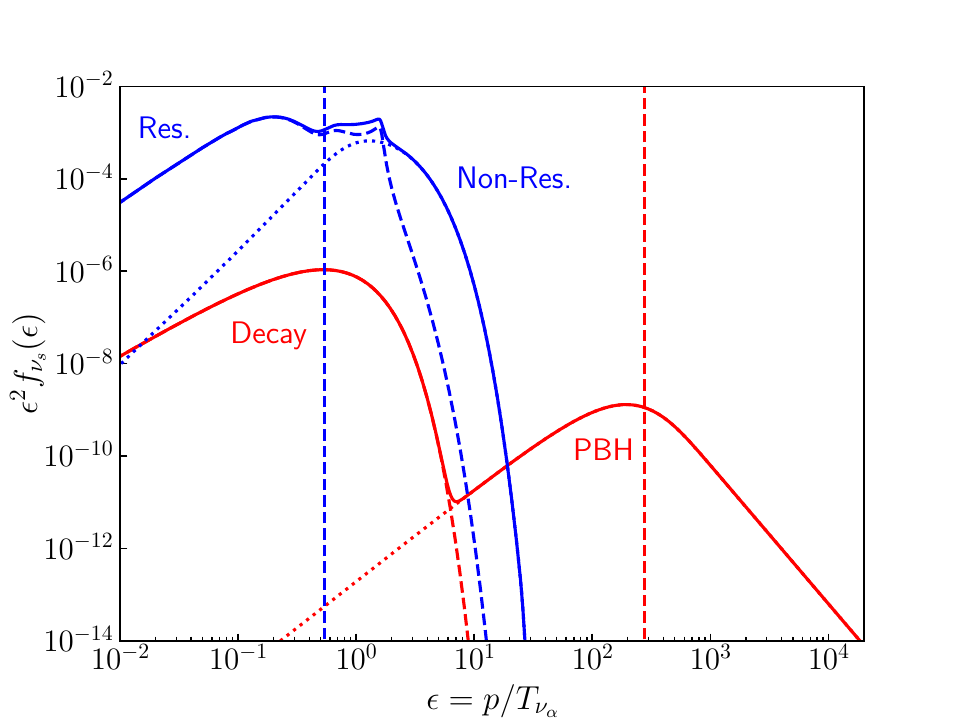}
    \caption{Representative momentum spectra of  
    two examples of two-population models.
    Shown in blue, resonant oscillation (dashed), non-resonant oscillation (dotted), and combined (solid) populations with $m_s = 7.8$ keV, $\sin^2 2\theta= 2.4 \times 10^{-11}$ and initial $L_e= 2.50 \times 10^{-3}$ (Model 2 of Tab.~\ref{tab:fs=1eg}). Shown in red, singlet Higgs decay (dashed), PBH neutrinogenesis (dotted), and combined (solid) with $m_s = 4$ MeV (Model 3 of Tab.~\ref{tab:pbhdecay}).  The vertical dashed blue and red lines show the average momentum corresponding to $m_{\rm therm}=5.7$ keV, separating cold and warm DM populations for $m_s=7.8$ keV and $m_s=4$ MeV, respectively.  
    }
    \label{fig:combined}
\end{figure}

In Tab.~\ref{tab:summary} we summarize four different $\nu_s$ production mechanisms considered in this work with corresponding typical population mass and momentum ranges. In Fig.~\ref{fig:combined} we highlights the distinct momentum distributions characteristic of each production mechanism, resulting in different implications for DM. Sterile neutrinos, like all DM candidates, are constrained by the requirement that their total contribution does not exceed the observed DM density with $f_{\rm s,DM} \leq 1$. In particular, sterile neutrinos produced via the DW mechanism alone cannot constitute all of DM due to combined warm DM and X-ray limits~\cite{Seljak:2006qw,Abazajian:2006yn,Viel:2006kd,Viel:2013fqw,Gelmini:2019wfp}.
However, sterile neutrinos produced through other mechanisms, including the multiple-population scenarios considered here, could still viably account for the entire DM abundance.

Warm DM, which becomes non-relativistic at $T \simeq \mathrm{keV}$, impacts structure formation on sub-galactic scales. Strong lensing observations of quasars statistically constrain the warm DM free-streaming length, disfavoring thermal relic DM masses below $m_{\rm therm} \sim 5~\mathrm{keV}$~\cite{Gilman:2019nap,Zelko:2022tgf}. In our analysis, we adopt comparable constraints from Lyman-$\alpha$ forest observations~\cite{Baur:2017stq,Garcia-Gallego:2025kiw}, specifically using the $2\sigma$ limits from Ref.~\cite{Garcia-Gallego:2025kiw}. These bounds on warm DM fraction can be directly applied to the two-population sterile neutrino scenarios considered here. The limits imply that a DM fermion with a thermal spectrum cannot constitute all of DM at the $2\sigma$ level if its mass is $m_{\rm therm} \leq 5.7$ keV. Furthermore, if $m_{\rm therm} \lesssim 1~\mathrm{keV}$, its contribution is restricted to $\lesssim 10\%$ of the total DM density. Heavier thermal fermions would have no discernible effect on structure at these scales and are classified as cold DM. 

These bounds on thermal relic DM can be translated into constraints on sterile neutrino parameters using the following relation, first introduced in~Ref.~\cite{Viel:2005qj,Baur:2017stq} and later refined in~Ref.~\cite{Gelmini:2019wfp} to account for the average scaled momentum $\langle \epsilon(t)\rangle$ at production 
\begin{equation}
\label{eq:lya}
m_s \simeq 4.46~\mathrm{keV} \left(\frac{\langle \epsilon(t) \rangle}{3.15}\right) \left(\frac{10.75}{g_*(t)}\right)^{1/3} \left(\frac{m_{\rm therm}}{1~\mathrm{keV}}\right)^{4/3} \left(\frac{0.12}{f_{\rm WDM} , \Omega_{\rm DM} h^2}\right)^{1/3}~.
\end{equation}
where $t$ is the production time, $f_{\rm WDM}$ is fraction of warm DM, and the dimensionless Hubble parameter $h\simeq0.67$.  
To account for production at different epochs with varying $g_{*}(t)$, we express the relation in terms of the entropy-scaled momentum $\epsilon_g = \epsilon(t)  g_{*}(t)^{-1/3}$ 
\begin{equation}
\label{eq:massrelation}
m_s \simeq 4.46~\mathrm{keV} \left(\frac{\langle \epsilon_g \rangle}{1.43}\right) \left(\frac{m_{\rm therm}}{1~\mathrm{keV}}\right)^{4/3} \left(\frac{0.12}{f_{\rm WDM} , \Omega_{\rm DM} h^2}\right)^{1/3}.
\end{equation} 
These relations allow us to map WDM constraints on $m_{\rm therm}$ into limits on sterile neutrino mass $m_s$ for a given production mechanism.
The average momentum $\langle \epsilon_g \rangle$ of the warm sterile neutrino population, combined with $m_s$, determines the effective $m_{\rm therm}$. By locating the corresponding point in the $(1/m_{\rm therm}, f_{\rm WDM})$ plane from Ref.~\cite{Garcia-Gallego:2025kiw}, we extract the maximum allowed fraction of warm DM, $f_{\rm s,lim}^{\rm warm}$. The sterile neutrino warm population must satisfy $f^{\rm warm}_{\rm s, DM} < f_{\rm s,lim}^{\rm warm}$. We also use these relations to determine the average momentum that for each given $m_s$ corresponds to $m_{\rm therm} = 5.7$ keV, shown in our figures as the boundary between cold and warm populations.

\begin{table}[t]
    \centering
    \begin{tabular}{|r|c|c|}
    \hline
    Production~~~~~~~ & Sterile neutrino & Sterile neutrino \\
    mechanism~~~~~~~ & mass $m_s$ range (keV) & momentum $\langle\epsilon\rangle_{\rm ref}$ range   \\
    \hline \hline
    non-resonant oscillations & $\simeq 1-10^3$ & 1.6-2.6 \\
    \hline
    resonant oscillations & ~$\simeq 10^{-3}-1$ & 0.1-1 \\
    \hline
    PBH neutrinogenesis & $\simeq 1-10^9$ & 5$\times 10^2$~-~$5\times10^{7}$ \\
    \hline
    heavy scalar decays & $\simeq 1-10^5$ & 0.9-2.1\\
    \hline
    \end{tabular}
    \caption{Summary of the four different considered sterile neutrino ($\nu_s$) production mechanisms with corresponding typical sterile neutrino mass and momentum ranges.}
    \label{tab:summary}
\end{table}

Sterile neutrinos as DM candidates can also be probed through their astrophysical radiative decays into photons, such as through the $\nu_s \rightarrow \nu_a \gamma$ channel. In the keV mass range, this decay would produce an X-ray line. Future X-ray experiments with improved sensitivity, such as a potential successor to NuSTAR~\cite{Neronov:2016wdd,Perez:2016tcq,Ng:2019gch}, covering the $\sim 10$–$200$ keV band~or the launched XRISM~satellite~\cite{XRISMScienceTeam:2020rvx}, sensitive to the $\sim 0.4$–15 keV range, could further test such scenario. The non-detection of such X-ray or $\gamma$-ray lines in current observations~\cite{Ng:2019gch,Perez:2016tcq,Neronov:2016wdd} places stringent upper limits on the sterile-active mixing angle $\sin^2 2\theta$ as a function of $m_s$.

\section{Conclusions}
\label{sec:conclusion}  

We have explored several scenarios in which a single sterile neutrino species can be produced in the early Universe with multiple, dynamically distinct relic populations, resulting in a complex, multi-peaked momentum spectrum. The phenomenology of such multi-population scenarios is qualitatively different from the typically considered single component sterile neutrino DM picture, with variety of implications including structure formation, X-ray observations  and small-scale cosmology.

Since sterile neutrinos mix with active neutrinos, some amount of production via active-sterile oscillations is inevitable in all scenarios. We have analyzed several conceptually distinct production pathways - resonant and non-resonant active-sterile flavor oscillations, gravitational production from PBH neutrinogenesis and heavy particle decays. These scenarios can combine to produce cold and warm sterile neutrino admixtures while remaining consistent with cosmological and astrophysical limits.

One general class of multi-population scenarios arises from active-sterile oscillations in the presence of a lepton asymmetry. A time-dependent lepton asymmetry $\mathcal{L}(T)$ initially produces a cold, resonantly enhanced sterile neutrino component, followed by a warmer, non-resonant contribution after the asymmetry is depleted. Solving the full Boltzmann equation while incorporating the temperature dependence of relativistic degrees of freedom, we find viable regions with sterile neutrino masses in the range $m_s \sim 4$–$10$ keV and mixing angles $10^{-12} \lesssim \sin^2 2\theta \lesssim 10^{-9}$. In these regions, a significant fraction of DM is produced via oscillations, yet the total abundance and spectral distribution satisfy both X-ray and Lyman-$\alpha$ constraints.

An entirely different mechanism involves sterile neutrino production from gravitational effects in PBH neutrinogenesis, where sterile neutrinos are generated through emission during Hawking evaporation. This production is independent of active-sterile mixing and can occur efficiently even for arbitrarily small mixing angles. We demonstrated that two different monochromatic in mass PBH populations can produce sterile neutrinos with two distinct spectral peaks, each tracking the evaporation temperature of the corresponding PBH mass. If the heavier PBHs temporarily dominate the Universe’s energy density, the earlier-emitted population is further redshifted, naturally generating cold and warm DM spectrum. Importantly, when PBH neutrinogenesis is combined with active-sterile oscillations, X-ray bounds constrain the oscillation-produced sterile neutrino fraction to be $\lesssim 10^{-10}$ of the total DM abundance, making it effectively negligible. Furthermore, PBH-dominated phases can source a stochastic gravitational wave background, providing an additional observational channel to probe this scenario.

We have also studied heavy particle decay as a production mechanism, particularly the decay of singlet scalars into sterile neutrinos. Scalar decays yield a cold, narrow momentum spectrum that is largely independent of the active-sterile mixing angle. When combined with a subdominant component from DW production, this scenario can simultaneously explain a potential X-ray line (e.g. the claimed but disputed 3.5 keV excess) while keeping the warm DM fraction below $\sim 10\%$, consistent with Lyman-$\alpha$ and strong lensing bounds. When accounting for all of the DM, the DW fraction is restricted by present observational limits to be negligible.

Additionally, we analyzed the combination of PBH neutrinogenesis with singlet scalar decay, which naturally produces widely separated momentum peaks. In this scenario, the PBH-produced component can constitute $\sim 10\%$ of the DM as warm dark matter, while the singlet decay-produced component forms the remaining $\sim 90\%$ as cold DM (see Fig.~\ref{fig:pbhdecay}). This mechanism offers another viable pathway to achieving mixed-phase DM while satisfying observational limits.

To systematically assess these scenarios, we developed semi-analytic tools combining numerical Boltzmann evolution with temperature-dependent degrees of freedom, analytic criteria for multi-modal spectra, and unified constraint mapping. Our approach translates Lyman-$\alpha$ limits on warm DM into direct bounds on the mass, momentum distribution, and fraction of each sterile neutrino population in the admixture. We identify broad parameter regions where the total DM relic abundance is achieved with a manifestly non-thermal, multi-component momentum distribution. These mixed-phase sterile neutrino models can simultaneously soften small-scale structure and evade current X-ray and lensing limits.

Looking ahead, this multi-population framework suggests various observational opportunities. Searches in upcoming X-ray experiments, such as recently launched XRISM and proposed Athena, will improve sensitivity to narrow emission lines from sterile neutrino decay. Surveys like DESI and Euclid will refine Lyman-$\alpha$ constraints on small-scale power suppression. Additionally, gravitational wave detectors may probe stochastic backgrounds from PBH evaporation, providing complementary evidence for PBH neutrinogenesis. As the next generation of experiments sharpen both small-scale cosmological probes and X-ray searches, the distinctive signatures of multi-population sterile neutrinos could soon be within experimental reach.

\section*{Acknowledgment}
\addcontentsline{toc}{section}{Acknowledgments}

This work was supported by World Premier International Research Center Initiative (WPI), MEXT, Japan. V.T. acknowledges support by the JSPS KAKENHI grant No. 23K13109. P.L. is supported by KIAS Individual Grant 6G097701. The work of GG was supported in part by the U.S. Department of Energy (DOE) Grant No. 
DE-SC0009937.

\appendix

\section{Analytic evaluation of non-resonant oscillation production}
\label{sec:nonres-theo}

Although in this work we solve the Boltzmann equation Eq.~\eqref{eq:boltz-eg-approx} numerically to obtain the sterile neutrino distribution function $f_{\nu_s}$, the non-resonant production can be estimated analytically.

In non-resonant production, the sterile neutrino production rate is sharply peaked around a temperature $T_{\rm max}$ (see Eq.~\eqref{eq:Tmax}). Therefore, it is a good approximation to fix the effective number of relativistic degrees of freedom at $g_{*} = g_{*}(T_{\rm max})$ and treat the temperature-scaled momentum at production as $\epsilon_{\rm prod} = \epsilon(T_{\rm max})$. In this approximation, we can use $\epsilon_{\rm prod}$ and $T$ directly instead of $\epsilon_g$ and $T_g$.
The temperature-time relation in Eq.~\eqref{eq:dTg1/3dt} then reduces to the standard form valid during radiation domination 
\begin{equation}
\label{eq:dT/dt}
\frac{dT}{dt}=-HT=-T^3\sqrt{\frac{8\pi^3g_*(T_{\rm max})}{90M_{\rm pl}^2}}~,
\end{equation} 
where $H$ is the Hubble expansion rate in a radiation-dominated Universe.
 
We may approximate the sterile neutrino distribution function $f_{\nu_s}(\epsilon_{\rm prod})$ by
\begin{equation}
\label{eq:fseps-dT}
    f_{\nu_s}(\epsilon_{\rm prod})\simeq -\int dT  \frac{1}{HT}\Gamma(\nu_\alpha\rightarrow\nu_s;\epsilon_{\rm prod},T)f_{\nu_\alpha}(\epsilon_{\rm prod})~,
\end{equation}
where $f_{\nu_\alpha}(\epsilon_{\rm prod}) = \left(\exp[\epsilon_{\rm prod}] + 1\right)^{-1}$ is the equilibrium Fermi-Dirac distribution, and the conversion rate is given by
\begin{eqnarray}
    \frac{1}{HT}\Gamma(\nu_\alpha\rightarrow\nu_s;\epsilon_{\rm prod},T)=\frac{1}{4HT} 
\left\{ \frac{\sin^22\theta\left[y_\alpha(T_{\rm max})G_F^2\epsilon_{\rm prod}T^5\right]}{\sin^22\theta+\left[\cos2\theta+2r_\alpha G^2_F\epsilon^2_{\rm prod} T^6/m_s^2\right]^2} \right\}~.
\end{eqnarray}  
In the absence of a large lepton asymmetry, the integrand in Eq.~\eqref{eq:fseps-dT} has the form $AT^2(1 + BT^6)^{-2}$, where $A$ and $B$ are constants. This integral can be solved exactly~\cite{Gelmini:2019wfp}, yielding
\begin{equation}
    \label{eq:fseps-nonres}
    f_{\nu_s}(\epsilon_{\rm prod})\simeq 9.25\times10^{-6}\left(\frac{\sin^22\theta}{10^{-10}}\right)\left(\frac{m_s}{\rm keV}\right)\left(\frac{g_*(T_{\rm max})}{30}\right)^{-1/2}f_{\nu_\alpha}(\epsilon_{\rm prod})~.
\end{equation}
For $\mathcal{L}_0 = 0$, sterile neutrinos and sterile antineutrinos obey the same Boltzmann equation, so the result in Eq.~\eqref{eq:fseps-nonres} applies to both $f_{\nu_s}$ and $f_{\bar{\nu}_s}$.

The number density change for any neutrino or antineutrino species with distribution function $f_\nu(\epsilon(t),t)$ at time $t$ is given by 
\begin{eqnarray}
    dn_{\nu}(\epsilon,t)=\frac{p^2f_{\nu}(p,t)dp}{2\pi^2}=\frac{T^3(t)}{2\pi^2}\epsilon^2f_\nu(\epsilon(t),t)d\epsilon~.
\end{eqnarray}
Using the approximation for $f_{\nu_s}(\epsilon)$ from Eq.~\eqref{eq:fseps-nonres}, the number density of sterile neutrinos produced around $t_{\rm prod} \simeq t(T_{\rm max})$ is  
\begin{eqnarray}
\label{eq:n_nu}
    n_{\nu_s}|_{t_{\rm prod}}\simeq 2\frac{T_{\rm max}^3}{2\pi^2}\int d\epsilon_{\rm prod}   \epsilon_{\rm prod}^2 f_{\nu_s}(\epsilon_{\rm prod})~.
\end{eqnarray}
where the factor of 2 accounts for both sterile neutrinos and antineutrinos.

After production, sterile neutrinos decouple from the thermal bath, and their number density redshifts with the expansion of the Universe. At any later time $t$, the number density becomes
\begin{equation}
    n_{\nu_s}(t)\simeq \frac{g_*(T)T^3(t)}{g_*(T_{\rm max})\pi^2}\int d\epsilon_{\rm prod}   \epsilon_{\rm prod}^2 f_{\nu_s}(\epsilon_{\rm prod})~.
\end{equation} 
In contrast, active neutrinos remain in thermal equilibrium until they decouple at $T(t_{\rm dc}) \simeq 3$ MeV. After decoupling, $e^+e^-$ annihilation heats the photon bath  resulting in a neutrino temperature $T_{\nu_\alpha} < T$ and a number density
\begin{eqnarray}
\label{eq:n_nu-alpha}
    n_{\nu_{\alpha}}(t)=\frac{T^3(t)}{\pi^2}\left(\frac{g_*(t)}{g_*(t_{\rm dc})} \right)^{1/3}\int d\epsilon_{\rm dc} \frac{\epsilon_{\rm dc}^2}{e^{\epsilon_{\rm dc}}+1}=\frac{3\zeta(3)}{2\pi^2}T_{\nu_\alpha}^3(t)~.
\end{eqnarray} 
For electron neutrinos decoupling at $T_{\rm dc} = 3$ MeV, we take $g_*(t_{\rm dc}) = 10.75$, giving a present-day electron neutrino number density of $n_{\nu_e,0} = 112~\mathrm{cm}^{-3}$.
The present-day sterile neutrino number density is then
\begin{eqnarray}
\label{eq:n_nu-s0}
n_{\nu_s,0}=\frac{2n_{\nu_e,0}}{3\zeta(3)}\left( 
\frac{10.75}{g_*(T_{\rm max})} \right) \int d\epsilon_{\rm prod}   \epsilon_{\rm prod}^2 f_{\nu_s}(\epsilon_{\rm prod})~.
\end{eqnarray}
Applying Eq.~\eqref{eq:fsdm} to the result in Eq.~\eqref{eq:n_nu-s0} yields the analytic estimate of the fraction of DM produced by the DW mechanism, as given in Eq.~\eqref{eq:fsdmnonresapprox}.
 
\section{Numerical Solution of Boltzmann Equation and Lepton Asymmetry}
\label{sec:num-cal}

For the numerical solution of the sterile neutrino Boltzmann equation, we discretize the production over small temperature intervals during which some relevant quantities remain approximately constant. We make use of the momentum variable
$\epsilon_g$ and generalized temperature $T_g$ to simplify calculations.

Consider the sterile neutrino production over an arbitrarily short $T_g$ interval $\delta T_g$, during which the effective number of relativistic degrees of freedom $g_*$ and the lepton asymmetry $L_\alpha$ can be treated as constant. In the interval $(T_g^{\rm prod}, T_g^{\rm prod} + \delta T_g)$, the generated sterile neutrino number density
at $T_g<T_g^{\rm prod}$ is
 
\begin{equation}
\label{eq:Deln-eg}
\begin{aligned}
    \Delta n_{\nu_s}(T_g^{\rm prod},\delta T_g;T_g)&\simeq \frac{T_g^3}{\pi^2}\int d\epsilon_g \epsilon_g^2\left.\int_{T_g^{\rm prod}-\delta T_g}^{T_g^{\rm prod}} dT_g'  \frac{df_{\nu_s}(\epsilon_g[g(T_g^{\rm prod})]^{1/3},T_g')}{dT_g'}\right|_{L_\alpha(T_g^{\rm prod})}\\
    &\simeq \frac{T_g^3}{\pi^2}\int d\epsilon_g \epsilon_g^2\Delta f_{\nu_s}(\epsilon_g[g(T_g^{\rm prod})]^{1/3};T_g^{\rm prod},\delta T_g)~.
    \end{aligned}
\end{equation}
The generated sterile neutrino follows the same equation with $\nu_s$ replaced by $\bar{\nu}_s$.
 
We also approximate the total change of lepton number during this period, 
by integrating Eq.~\eqref{eq:dLdtv2} with a constant lepton number $L_{\alpha}(T_g^{\rm prod})$ evaluated at $T_g=T_g^{\rm prod}$. We then obtain
\begin{equation}
\label{eq:DelL-dc}
    \Delta L_\alpha(T_g^{\rm prod},\delta T_g)\simeq -\frac{[g_*(T_g^{\rm prod})](T_g^{\rm prod})^3}{2\pi^2n_\gamma(T_g^{\rm prod})}\int_{T_g^{\rm prod}-\delta T_g}^{T_g^{\rm prod}} dT_g\int d\epsilon_g  \epsilon_g^2\left.\left[ \frac{df_{\nu_s}}{dT_g}- \frac{df_{\bar{\nu}_s}}{dT_g}\right]\right|_{L_\alpha(T_g^{\rm prod})}~.
\end{equation}
The fact that $\epsilon_g$ and $T_g$ are independent allows the $T_g$ integral to be computed first. Eq.~\eqref{eq:DelL-dc} simplifies to
\begin{equation}
\label{eq:DelL-deln}
    \Delta L_\alpha(T_g^{\rm prod},\delta T_g)\simeq -\frac{g_*(T_g^{\rm prod})}{2n_\gamma(T_g^{\rm prod})}[\Delta n_{\nu_s}(T_g^{\rm prod},\delta T_g)-\Delta n_{\bar{\nu}_s}(T_g^{\rm prod},\delta T_g)]~.
\end{equation}
Here, $\Delta n_{\nu_s}$ and $\Delta n_{\bar{\nu}_s}$ are the sterile neutrino and antineutrino number density changes during the interval.
The corresponding change $\Delta f_{\nu_s}$ in the momentum distribution during this period, appearing in Eqs.~\eqref{eq:Deln-eg} and~\eqref{eq:DelL-dc}, can be calculated by numerically solving the Boltzmann equation as given in Eq.~\eqref{eq:boltz-eg-approx}. Using the computed $n_{\gamma}(T_g^{\rm prod})$ and Eq.~\eqref{eq:Deln-eg} for $\Delta n_{\nu_s}(T_g^{\rm prod}, \delta T_g)$ and $\Delta n_{\bar{\nu}_s}(T_g^{\rm prod}, \delta T_g)$, the change in lepton number from Eq.~\eqref{eq:DelL-deln} can be evaluated for any $T_g^{\rm prod}$ and arbitrarily small $\delta T_g$.

In our numerical calculations, we discretize the $T_g$ domain into small temperature intervals $\Delta T_g$, chosen such that $g_{*}(T_g - \Delta T_g) \simeq g_{*}(T_g)$. The temperature dependence of $g_{*}(T_g)$ is shown in Fig.~\ref{fig:gstar}. The initial temperature $T_g^{i}$ is selected such that the thermal potential $V_T$ is large enough to suppress production, ensuring $\dot{L}_e(T_g^{i}) \simeq 0$. This condition guarantees that $f_{\nu_s}(T_g^{i}) \simeq 0$ and $L_e(T_g^{i}) \simeq L_{e,0}$, allowing us to neglect the history before $T_g^{i}$.
The final temperature is set to $T_g^f \simeq 3$ MeV, corresponding approximately to the electron neutrino decoupling temperature.
Integrating Eq.~\eqref{eq:boltz-eg-approx} over the $j$-th temperature interval yields 
\begin{eqnarray}
\label{deltaf-appB}
    \Delta f_{\nu_s}^j&=&-\int_{T_{g}^j-\Delta T^j_{g}}^{T_g^j} \frac{dT_g'}{H(T_g')T_g'} \Gamma(\nu_\alpha\rightarrow\nu_s;\epsilon_g,T_g')f_{\nu_\alpha}(\epsilon_g,T_g')\\
    &\simeq&-\frac{1}{H(T_g^j)T_g^j}\Gamma(\nu_\alpha\rightarrow\nu_s;\epsilon_g,T_g^j)f_{\nu_\alpha}(\epsilon_g,T_g^j)\Delta T_g^j\notag~.
\end{eqnarray} 
At high temperatures $T_g \gtrsim \mathcal{O}(\mathrm{GeV})$, the conversion rate is slow, and relatively large step sizes $\Delta T_g \sim \mathcal{O}(\mathrm{MeV})$ provide sufficient accuracy. However, near the resonance at $T \simeq T_{\rm end}$, the production sharply peaks, and much smaller step sizes $\Delta T_g \sim \mathcal{O}(\mathrm{keV})$ are required to resolve the rapid changes in the sterile neutrino distribution and the lepton asymmetry. Failure to do so can lead to numerical overshooting of the lepton asymmetry, potentially resulting in an unphysical negative lepton number.
By implementing suitable adaptive step sizes near the resonance region, we ensure numerical stability and find that the net driving lepton asymmetry remains positive for $L_{\mu,\tau} = 0$. This behavior is expected from the form of the conversion rates in Eqs.~\eqref{eq:totalrate-nu} and~\eqref{eq:totalrate-nubar}, which approach equality as $\mathcal{L} \rightarrow 0$, naturally leading to $\dot{\mathcal{L}} \rightarrow 0$ via Eq.~\eqref{eq:dLdtv2}.

With an appropriately chosen temperature sample set ${T_g^j}$, we calculate $\Delta f_{\nu_s}^j$ in each interval using Eq.~\eqref{deltaf-appB}. These values are then used in Eqs.~\eqref{eq:Deln-eg} and~\eqref{eq:DelL-deln} to compute the corresponding changes in the number density $\Delta n_{\nu_s}^j$ and lepton asymmetry $\Delta L_{\alpha}^j$.
By summing over all steps, the final sterile neutrino distribution at the reference temperature $T_{\rm ref}$ is given by
\begin{eqnarray}
\label{f-as-sun-of-deltaf-appB}
  f_{\nu_s}(T_{\rm ref})\simeq\sum_j \Delta f^j_{\nu_s}(\epsilon_{\rm ref})~.
\end{eqnarray}
Similarly, the total production is
\begin{equation}
    n_{\nu_s}(T_{\rm ref})\simeq\sum_i \Delta n_{\nu_s}^j~,
\end{equation}
and the total lepton number change is then
\begin{equation}
    \Delta L_\alpha(T_{\rm ref})\simeq-\frac{1}{2n_\gamma(T_{\rm ref})}[n_{\nu_s}(T_{\rm ref})-n_{\bar{\nu}_s}(T_{\rm ref})]~.
\end{equation}

Previously, in Ref.~\cite{Kishimoto_2008}  the Boltzmann and lepton number equations were solved assuming a constant effective number of relativistic degrees of freedom, $g_* \simeq 61.75$. In contrast, Ref.~\cite{Kasai:2024diy} solved both equations with a time-dependent $g_*(t)$, but using more complex numerical procedures.

In this work, we present a new and simpler method to systematically solve the Boltzmann equation alongside the time evolution of the lepton asymmetry while fully accounting for the temperature(time) dependence of $g_*$. By expressing the Boltzmann equation in terms of the $(\epsilon_g, T_g)$ variables defined in this paper, we eliminate additional derivative-like terms. This formulation allows for straightforward integration over $T_g$, providing an efficient and accurate solution to the coupled system.

\bibliography{refs}
\addcontentsline{toc}{section}{Bibliography}
\bibliographystyle{JHEP}

\end{document}